\documentclass{article}
\usepackage[utf8]{inputenc}
\usepackage{jcappub}

\usepackage{amsmath,amsfonts,amssymb}
\usepackage{mathrsfs}
\usepackage{graphicx}
\usepackage[english]{babel} 
\usepackage{hyperref} 
\usepackage[dvipsnames]{xcolor}
\usepackage{adjustbox}
\usepackage[normalem]{ulem}
\usepackage[capitalize]{cleveref}
\usepackage{csquotes}

\def\mpl{M_{\rm Pl}}

\title{News from the Swampland -- Constraining string theory with astrophysics and cosmology}
\date{July 2023}

\author[1]{Nils Sch\"oneberg,}
\affiliation[1]{Institut de Ci\`encies del Cosmos, Universitat de Barcelona, Mart\'{\i} i Franqu\`es 1, Barcelona 08028, Spain}
\author[2]{L\'eo Vacher,}
\affiliation[2]{Institut de Recherche en Astrophysique et Planétologie, CNRS/CNES, 9 avenue du Colonel Roche BP, Toulouse, France\\}
\affiliation[2]{Universit\'e de Toulouse UPS, 118 route de Narbonne, Toulouse, France\\}
\author[3,4,5]{J. D. F. Dias,}
\author[3,4]{Martim M. C. D. Carvalho,}
\author[3,5]{C. J. A. P. Martins}
\affiliation[3]{Centro de Astrof\'{\i}sica da Universidade do Porto, Rua das Estrelas 4150-762 Porto, Portugal \\}
\affiliation[4]{Faculdade de Ci\^encias, Universidade do Porto, Rua Campo Alegre 4169-007, Porto, Portugal\\}
\affiliation[5]{Instituto de Astrof\'isica e Ci\^encias do Espa\c co, CAUP, Universidade do Porto, Rua das Estrelas, 4150-762, Porto, Portugal}

\abstract{Our current best guess for a unified theory of gravitation and quantum field theory (string theory) generically predicts a set of requirements for a consistently quantized theory, the Swampland criteria. Refined versions of these criteria have recently been shown to be in mild tension with cosmological observations. We summarize the status of the current impact of and constraints on the Swampland conjectures from cosmology, and subject a variety of dark energy quintessence models to recently released cosmological datasets. We find that instead of tightening the tension, the new data allows for slightly more freedom in the Swampland criteria. We further demonstrate that if there is no theoretical argument made to prevent interactions of the moduli fields with the electromagnetic sector, a novel fine-tuning argument arises from the extremely tight current constraints on such interactions. Finally, we conclude with a cautionary tale on model-independent reconstructions of the Swampland criteria from expansion rate data.}

\begin{document}
\emailAdd{nils.science@gmail.com}

\maketitle

\section{Introduction}

In the search for the fundamental physics underlying our universe the prediction of additional dynamical scalar fields is ubiquitous. They appear to be necessary components for most unifying frameworks beyond our contemporary standard models of particle physics and gravity, such as string theory. The spectrum of possible scalar field theories viable at the classical level is extremely large. In string theory, the different choices of compactification schemes for the extra spatial dimensions of space-time give rise to a vast  number of possible solutions ($\gtrsim 10^{755}$ \cite{Halverson2017}), each featuring different scalar degrees of freedom associated to the compactified metric: the moduli fields. It is therefore a critical matter to understand and classify these solutions. The so-called Swampland program aims to find criteria on which to distinguish viable theories, part of the so-called Landscape, from the unacceptable ones, which constitute the Swampland. These criteria are based on the possibility to consistently quantize the theory, and as such related to UV completeness \cite{Vafa2005}, and have been confirmed even in exotic corners of the string theory space \cite{Brinkmann:2021czt}.
For a complete review see \cite{Palti2019,vanBeest:2021lhn}, and for critical discussions about the validity of the Swampland criteria see \cite{Akrami:2018ylq,Dasgupta:2018rtp,Kallosh:2019axr,Blumenhagen:2019qcg,Blumenhagen:2019kqm,Banerjee:2021yrb}.

Scalar fields also underpin much of cosmology (such as in inflation, late-time dark energy, or possibly early dark energy), arising as foundational Lagrangian building blocks in a multitude of extensions (and serving also as useful tools to approximate more complicated models). As such, it is fundamentally important to ask how they can be constructed in a consistent unified theory. Indeed, Swampland conjectures have been extensively confronted with inflation models, showing significant tensions and providing strong restrictions on the allowed parameter space of the simplest class of models: those involving a single field\footnote{As we will see later, the de-Sitter Swampland criteria essentially requires the slow roll conditions to be explicitly broken, making single field inflation rather difficult to construct consistently.} \cite{Agrawal2018,Scalisi:2018eaz,Blumenhagen2018,Scalisi:2019gfv,Fukuda2019,NooriGashti2022,Freigang:2023ogu}. However, hope is not lost for inflationary scenarios remaining valid even when Swampland criteria are imposed, as a large variety of proposals has been put forth to rescue the inflationary paradigm. We list a small excerpt of such proposals below:
\begin{itemize}
    \item In the context of multi-field inflation see e.g. \cite{Achucarro2019,Garg2019, Andriot2019,Ben-Dayan2019,Bravo2020,Bertolami2022}.
    \item In the context of warm inflation see e.g. \cite{Das2019,Gungor2011,Kamali2020,Das2020,Mohammadi2020,Brandenberger2020,Trivedi2022,dosSantos2022}.
    \item In the context of eternal inflation see e.g. \cite{Matsui2019,Wang2020,Kinney2019,Lin2020}.
    \item In the context of natural or Starobinsky inflation see e.g.  \cite{Fukuda2019, Osses2021}.
    \item In the context of other inflationary paradigms see e.g.  \cite{Kinney2019-II,Holman2019,Bjorkmo2019,Geng2020,Adhikari2020,Storm2020,Anchordoqui2021}.
    \item In the context of modified gravity theories, see e.g. \cite{Artymowski2019, Heisenberg2019-II, Benetti2019, Sadeghi2019,Oikonomou2022, Sadeghi2021,NooriGashti:2022siq,Sadeghi:2021cfs,Sadeghi:2022tzd,Gashti:2022hey}.
    \item In the context of avoiding Swampland conjectures through other means, see e.g. \cite{Brahma2019-II,Ashoorioon2019} for excited initial states, \cite{Trivedi2022-II} for Lorentz violation, and \cite{Brahma2020} for a tunneling wavefunction.
\end{itemize}
See also \cite{Kamali2020-II} for a discussion of re-heating in the context of the Swampland conjectures. Given this vast space of reconciling inflation with the Swampland conjectures, it is only natural to ask what other parts of cosmology feature scalar fields and how they are impacted by the conjectures.

%It has been argued that these constraints can be circumvented by multi-field inflation models, which, assuming the validity of the conjectures, could thereby be favored over single field models \cite{Achucarro2019}, but this point remains subject to discussion \cite{Ashoorioon2019}.

The widely accepted $\Lambda$CDM cosmological model is built on the presence of a cosmological constant $\Lambda$ whose origin is not trivial to explain, especially as computations of the vacuum energy in the standard model typically result in expectations many orders of magnitude higher \cite{Weinberg1989}. As such, in most high energy extensions of the standard model the cosmological constant is abandoned in favor of a dynamical dark energy.
Moreover, in many realizations of string theory the dark energy -- responsible for the late time acceleration of the universe -- must necessarily be promoted to a dynamical field, subject to the same Swampland criteria. Indeed, due to the challenges of constructing a de Sitter vacuum in string theory, it is possible to motivate Swampland conjectures arguing that any such vacuum is not part of the Landscape. In turn this implies that -- in a consistent theory of quantum gravity -- dark energy must be dynamical at some level \cite{Obied2018}. Therefore the Swampland conjectures can also be used to investigate the nature of dark energy. While our current $\Lambda$CDM model is built on the assumption that dark energy is sourced by a single cosmological constant, $\Lambda$, emerging from general relativity, it seems extremely difficult to build such a state from string theoretical considerations.

Quintessence provides the simplest form of dynamical dark energy in the form of a scalar field driving the late time acceleration of our universe's expansion \cite{Caldwell1998}. It is clear that if one accepts the Swampland conjectures, many of the arguments and constraints applying in the inflationary context will also apply here, and therefore the Swampland conjectures also allow to put stringent restrictions on the possibility for a scalar field to source dark energy. In the context of quintessence, the Swampland conjectures have been invoked in \cite{Akrami:2018ylq,Heisenberg2018,Heisenberg2019,Wang2018,Raveri2019,Colgain2019,Trivedi2020}, and in other late-time model-specific contexts in \cite{Cai2019,Brahma2019,vandeBruck2019,Mukhopadhyay2019,Anchordoqui2020,Montefalcone2020,Oikonomou2021}. Instead, analyses of the implications of the Swampland conjectures have also been made for the low-redshift universe in a more model-independent way in \cite{Yang2020,Elizalde2019,Elizalde2021,Banerjee2021,Arjona2021,Heisenberg2021}. The general conclusion of these works is a mild but not insurmountable tension between late time observations of dark energy and the Swampland criteria.

In the present work we aim to give a compressed overview of the Swampland criteria in the context of cosmology, with a particular focus on its applications to late-time cosmology and single-field quintessence models. In particular, we are going to outline that the present tension persists in a weakened form with new data, both for model-specific and model-independent analyses. We are also going to present novel fine-tuning arguments about possible interactions of the quintessence fields with other standard model fields. These arguments show that if the interactions are not forbidden for some deeper physical reason, current local and astrophysical tests on the stability of the fine structure constant $\alpha$ are significantly more stringent than analogous constraints on dark energy, and therefore have a significant impact on the consequences of the conjectures.

In \cref{sec:theory} we shortly review the theoretical foundations of the Swampland program with a special focus on the applicability to quintessence. We then present our methodology and data in \cref{sec:methodology}, while our results will are shown in \cref{sec:results}. We conclude in \cref{sec:conclusions}.

\section{Theory}\label{sec:theory}

In this section we briefly review some theoretical foundations, and establish our notation. The interested reader is encouraged to look into the reviews cited below for more information beyond the succinct summary presented here. In \cref{ssec:swampland} we summarize the state of the Swampland conjectures in terms of cosmology, and in \cref{ssec:quintessence} we briefly summarize the quintessence model and its relation to the Swampland conjectures.

\subsection{The Swampland}\label{ssec:swampland}
Here we only provide a short succinct summary of the state of the Swampland program, for more details we refer the reader to \cite{Akrami:2018ylq,Palti2019,vanBeest:2021lhn}.

At the present stage, the Swampland program consists of a set of conjectures, generally based on attempts to construct stable quantum theories of gravity. Typically, one encounters such conditions when seeking for viable vacuums of string theory, when making considerations regarding black holes, or when assessing the possibility to recover effective field theories emerging from underlying micro-physics \cite{Palti2019}. 

Perhaps one of the most fundamental conjectures is the distance conjecture. In its fundamental form it simply states the necessity of an infinite tower of mass states (related to the expectation values of the underlying fields), which occurs somewhere at an infinite distance within the (non-compact) manifold spanned by the fields (the moduli space) \cite{Ooguri2007}. However, this distance conjecture has been recently refined in \cite{Baume2016,Klaewer2017}. In its refined form, the conjecture argues that the infinite tower of light states already appears for trans-Planckian field variations and even in the case of moduli fields subject to a potential. Invoking an upper bound on the allowed number of light states\footnote{This is not as trivial as it might seem, as these mass states could not only be particles but also extended objects such as cosmic strings. We are going to remain agnostic and follow the literature in this regard, and formulate the conjecture as a restriction of the possible low-energy theory.} (such as the Bousso bound \cite{Bousso1999}, which extends the Bekenstein bound on space-time entropy \cite{Bekenstein1972,Bousso2003}, or alternatively arguments about the breakdown of the effective low-Energy field theory \cite{vanBeest:2021lhn}), one can convert this statement into the requirement that trans-Planckian excursions can not be allowed for any fields present in the cosmological evolution. We will refer to this specialized Swampland distance conjecture simply as the \textbf{distance conjecture}:
\begin{equation}\label{eq:swampland_distance}
    |\Delta \phi|/M_\mathrm{pl} < \mathcal{O}(1)~,
\end{equation}
with $\mpl$ being the reduced Planck mass\footnote{We work in natural units ($c=\hbar=1$).}, $\mpl^{-2} = 8\pi G$. It should be kept in mind that there is a small degree of variation of the constraint in the values for many string theory realizations, so the exclusion of a unity-order value is not strictly a large problem.

Due to the difficulties of consistently constructing the meta-stable de-Sitter vacua at the heart of cosmology, \cite{Ooguri2019}\footnote{The first swampland criterion had already been proposed in for example \cite{Obied2018}, though the connection to the distance conjecture has been clarified in \cite{Ooguri2019}.} has further proposed a requirement on possible field potentials of theories in the Landscape, given by
\begin{equation}\label{eq:swampland_desitter}
\textbf{either}\quad\mpl\frac{\left|V'\right|}{V} \gtrsim \mathcal{O}(1) \qquad \textbf{or}\quad -\mpl^2\frac{V''}{V} \gtrsim \mathcal{O}(1)~,
\end{equation}
where a prime denotes the derivative with respect to the field. We are going to denote these two alternative de-Sitter constraints as the \textbf{first Swampland criterion} and \textbf{second Swampland criterion}, respectively. Note that a combined further refinement has been put forth in \cite{Andriot2019-II}. Finally, in order to simplify notation, we introduce 
\begin{equation}\label{eq:def_lambda}
    \lambda(z) = \mpl |V'(z)/V(z)|~,\qquad \text{and}\qquad g(z) = -\mpl^2 V''(z)/V(z)~.
\end{equation} 
Any one of these two critical functions (of which we have made the redshift dependence explicit) needs to be positive and at least of order unity for the de-Sitter constraints to hold. There are also a few criteria beyond these most well-known ones, such as excellently summarized in \cite{Palti2019}, but we are not going to use those in the context of this work.

\subsection{Single-field Quintessence}\label{ssec:quintessence}
Quintessence is a framework in which dark energy is a dynamical degree of freedom, responsible for the accelerated expansion seen in recent cosmological history. It is a simple modification of $\Lambda$CDM, where the cosmological constant is replaced by one or more additional scalar fields minimally coupled to gravity as:
\begin{equation}
    S = \int d^4x \sqrt{-g} \cdot \Big\{\mpl^2\frac{R}{2} - \frac{1}{2}g^{\mu\nu}G_{ij}\partial_\mu\phi^i\partial_\nu\phi^j - V(\phi) + ... \Big\}~,
\label{eq:full-action}
\end{equation}
where $G_{ij}$ is a metric over field space, which reduces to 1 in the single-field case, $g_{\mu\nu}$ is the usual metric and $g$ its determinant. Assuming a single field, it is quite simple to find the Klein-Gordon equation of motion at the background level
\begin{equation}\label{eq:cqf_kg}
    \ddot{\phi} + 3 H \dot{\phi} + V'(\phi) = 0 ~,
\end{equation}
and at the level of the field perturbation $\delta \phi$ with
\begin{equation}\label{eq:cqf_perturbed_kg}
    \ddot{\delta \phi} + 3 H \dot{\delta \phi} + \dot{h} \dot{\phi}/2 + \left[k^2/a^2 + V''(\phi)\right] \delta\phi = 0~,
\end{equation}
in the comoving synchronous gauge (for the Newtonian gauge, see \cite[Eq.~4.3]{Archidiacono:2022iuu}). Here $V'(\phi)$ and $V''(\phi)$ denote the first and second derivatives of the potential $V(\phi)$ with respect to the field $\phi$. The field density and pressure are given by
\begin{equation}\label{eq:field_density_pressure}
    \rho_\phi = \frac{1}{2} \dot{\phi}^2 + V \qquad \mathrm{and} \qquad P_\phi = \frac{1}{2} \dot{\phi}^2 - V~.
\end{equation}
These are also related by the trivial energy conservation $\dot{\rho_\phi} = -3H(\rho_\phi+P_\phi) = -3H \dot{\phi}^2$ that is directly equivalent to \cref{eq:cqf_kg}. If the field is dominated by the potential contribution its equation of state $w_\phi = P_\phi/\rho_\phi$ tends towards $-1$, while in the kinetically dominated regime it tends towards $+1$.

\noindent For the Friedmann equation one can simply add the field energy density to the other densities:
\begin{equation}\label{eq:cqf_friedmann}
    H^2 = \frac{8\pi G}{3}\rho_\mathrm{tot} = \frac{8\pi G}{3} \Big[ \rho_a + \frac{1}{2}\dot{\phi}^2 + V(\phi) \Big]~,
\end{equation}
where $\rho_a$ denotes all other components contributing to the energy density (such as radiation, cold dark matter, baryons, neutrinos, ...). The energy conservation of the total content of the universe implies immediately that
\begin{equation}\label{eq:cont}
    \dot{\rho}_\mathrm{tot} = -3H(\rho_\mathrm{tot} + P_\mathrm{tot}) = -3H\rho_\mathrm{tot}(1+w_\mathrm{eff})~,
\end{equation}
where $w_\mathrm{eff} = P_\mathrm{tot}/\rho_\mathrm{tot} = (P_a + P_\phi)/(\rho_a + \rho_\phi) = r_a w_a + r_\phi w_\phi$, with $r_x = \rho_x/\rho_\mathrm{tot}$ and $w_x=P_x/\rho_x$\,. In order to check what conditions the field must satisfy in order to be responsible for the universe's acceleration, we can differentiate the Friedmann equation with respect to time, substituting $\dot{\rho}_\mathrm{tot}$ for \cref{eq:cont} and divide by $2H$, obtaining the well known second Friedmann equation
\begin{equation}
    \dot{H} = -\frac{3}{2} H^2(1+w_\mathrm{eff})~,
\end{equation}
which can be explicitly written to isolate the acceleration of the universe
\begin{equation}
    \ddot{a} = -\frac{1}{2}aH^2 (1 + 3w_\mathrm{eff})~.
\end{equation}
Clearly, to allow for the accelerated expansion of the universe observed at late times, we need $w_\mathrm{eff}< -1/3$ to achieve $\ddot{a} > 0$. This immediately allows us to put a constraint on the equation of state of the scalar field such that $w_\phi < (-1/3 - r_a w_a)/r_\phi$, with $r_a>0$ and $w_a \approx 0$ at low redshift ($z<5$), where the dark energy can become important. As such, we are left with approximately $w_\phi < -1/(3 r_\phi)$, requiring at least a mild domination of the potential energy of the field with respect to the kinetic energy. The underlying problem of fulfilling the Swampland criteria in this regime is that a large $V'(\phi)$ will -- through \cref{eq:cqf_kg} -- typically quickly accelerate the field and thus make it hard for the overall potential to dominate. 

\begin{figure}
    \centering
    \includegraphics[width=\textwidth]{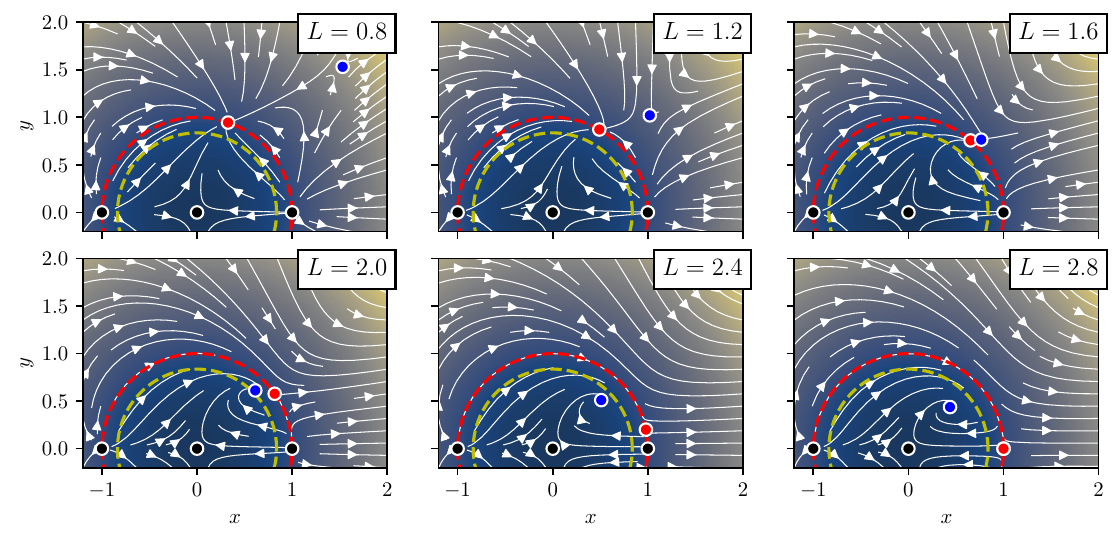}
    \caption{Differential equation flows in the $(x,y)$ plane for various values of $L=-\mpl V'/V$. The arrows show the directions of $(\dot{x},\dot{y})$. The red dashed line shows $x^2+y^2=1$, while the yellow dashed line shows $x^2+y^2=0.7$. The colored dots denote the critical points (where the gradient is zero). The trivial critical points at $y=0$ are colored in black. For $L<\sqrt{3}\simeq 1.73$ the blue fixed point is repulsive and further from the origin than the red fixed point at $x^2+y^2=1$. For $L>\sqrt{3}$ the blue point is closer to the origin and becomes attractive, while the red point becomes repulsive. Only as long as $L \lesssim \sqrt{3/(1-\Omega_m)} \approx 2.1$ is the blue point at $x^2+y^2 > 1-\Omega_m$ (outside the yellow circle). At large $L>\sqrt{6}\simeq 2.45$ the red point joins the trivial critical point at $(x,y)=(1,0)$.}
    \label{fig:lambda_flows}
\end{figure}

\noindent In order to analyze the dynamics of the quintessence model, it is often useful to reparameterize it through \cite{Tsujikawa2013}
\begin{equation}
    x=\frac{\dot{\phi}}{\sqrt{6}\mpl H}~, \qquad y=\frac{\sqrt{V}}{\sqrt{3}\mpl H}~.
\end{equation}
Using these variables, one simply has $\rho_\phi/(3\mpl^2 H^2) = \rho_\phi/\rho_\mathrm{tot} = x^2+y^2$, so iso-density contours represent a simple circular arc. The field and Friedmann equations then give rise to non-linear first-order differential equations for $x$ and $y$
\begin{align}
    \frac{\partial x}{\partial \ln a} = -3x + \frac{\sqrt{6}}{2} L \cdot  y^2 - x \frac{\partial \ln H}{\partial \ln a}~,\label{eq:trajectory_x} \\
    \frac{\partial y}{\partial \ln a} =  -\frac{\sqrt{6}}{2} L \cdot  x y - y \frac{\partial \ln H}{\partial \ln a}~. \label{eq:trajectory_y}\\
\end{align}
where $L = -\mpl V'/V$ (such that from \cref{eq:def_lambda} we have $\lambda(z) = |L|$). The flow of the trajectories in this $(x,y)$ space can be analyzed using a dynamical system approach. We show the possible trajectories in \cref{fig:lambda_flows} for a few values of $L$, with the red iso-density contour marking complete dark energy domination ($x^2+y^2=\rho_\phi/\rho_\mathrm{tot} = 1$), and the yellow iso-density contour marking the observation of dark energy density in the late universe from supernovae data, at around ($x^2+y^2 \simeq 0.7$). The two non-trivial fixed points are $(x,y) = (L/\sqrt{6},\sqrt{1-L^2/6})$ (red dot) and\footnote{The true fixed point is at $(x,y)=(\sqrt{3/2}/L \cdot (1+w_a), \sqrt{3/2}/L \cdot \sqrt{1-w_a^2})$. In order to simplify the discussion we explicitly set $w_a=0$ (which is a very good approximation at the redshifts of interest). In the full parameter space the point approaches the $x^2+y^2=1$ boundary only as $L^2 \to 3 (1+w_a)$, but otherwise the discussion remains the same.} $(x,y) = (\sqrt{3/2}/L,\sqrt{3/2}/L)$ (blue dot). The red fixed point at $x^2+y^2=1$ is the growing attractor solution as long as the blue solution does not reach the boundary, i.e. for $L\leq\sqrt{3}\simeq 1.73$. This means that for $L^2 > 3$ the universe cannot ever be fully dominated by dark energy, even in the infinite future. However, this is technically not even required by the current observations, which only set a weaker condition that the current amount of cosmic acceleration has to be explained by some dark energy, which gives the weaker condition of $x^2+y^2 \simeq (1-\Omega_m) \approx 0.7$. This condition can be fulfilled for even higher values of~$L$. At the point where $L =\sqrt{3}$ the two non-tirival fixed points intersect at $(x,y)=(1/\sqrt{2},1/\sqrt{2})$, after which they exchange their attractive/repulsive behavior. The attractive point is now the blue one, which for a given $L$ can reach up to $x^2+y^2 = 3/L^2$. This means that for the aforementioned reason we instead have the weaker condition $L \lesssim 2.1$. 

Note that the system of \cref{eq:trajectory_x,eq:trajectory_y} obeys the symmetry $(x,y,L) \to (-x,-y,-L)$, so the condition can be rephrased equivalently as $\lambda(z) \lesssim 2.1$. This implies that the observed acceleration of the universe can only be realized as long as $\lambda(z) \lesssim 2.1$, whereas the first Swampland criterion requires $\lambda(z) \gtrsim \mathcal{O}(1)$. It is this inherent tension that makes the observed acceleration such a potent probe of the Swampland criteria.

\noindent The argument for the second Swampland criterion can be built upon the same basis, but is a bit more subtle. Here one can use
\begin{equation}
    \mathrm{d}L/\mathrm{d}\ln a = \sqrt{6}\mpl \cdot x \cdot (L^2+g)~,
\end{equation}
where $g(z)=-\mpl V''/V$ as in \cref{eq:def_lambda}. If $g \gg 1$ we find that $L$ is quickly accelerating, which is a self-sustaining process\footnote{Of course it is only self-sustaining since $g>0$ in this case. In the exponential potential case, for example, we have $g = - L^2 < 0$, thus giving $\mathrm{d}L/\mathrm{d}\ln a = 0$ as expected.} as evident from the equation. In this case the growing $L$ will soon violate $\lambda(z) < 2.1$ required for the accelerating universe at the attractor.

Overall, we can summarize that the fulfillment of either Swampland criterion is in mild tension with the observation of an accelerating universe in the quintessence model. Exactly how strong this tension is will be investigated in \cref{sec:results}, but even the simple analytical arguments performed here already demonstrate that the observation of the universe's expansion history can necessarily place tight constraints on the Swampland criteria.
%At the dark energy dominated attractor solution $(x,y) = (\lambda/\sqrt{6},\sqrt{1-\lambda^2/6})$, we have $w_\mathrm{eff} = w_\phi = \lambda^2/3 - 1$, such that $\lambda^2 < 2$ correspond to accelerating solutions. This limit is well within the range of $\lambda$ below criticality, because $0 \leq w_m \leq 1$. Thus, cosmologies with $2 \leq \lambda < 3(1+w_m)$ are viable, but non-accelerating, solutions.

%For a review in the context of late time cosmology see e.g. \cite{Raveri2019}.

%Refined de-Sitter conjecture \cite{Obied2018,Ooguri2019,Garg2019}:
%\begin{equation}
%    \lambda := \mpl\frac{\left|V'\right|}{V} \gtrsim \mathcal{O}(1) \qquad {\rm or} \qquad c^2:=-\mpl^2\frac{V''}{V} \gtrsim \mathcal{O}(1)
%\end{equation}
%Distance conjecture
%\begin{equation}
%    \frac{\Delta \phi}{\mpl} \lesssim  \mathcal{O}(1)
%\end{equation}

\subsection{Coupling to electromagnetism}
In the most general setting, one expects new scalar fields $\phi$ present in the theory to be coupled to all the sectors of the standard model, unless some symmetry principle prevents this. There is no strong reason a priori why a UV-complete theory formulated within string theory should decouple such scalar fields from other fields in the low-energy limit applicable to most of cosmology. This, in turn, causes a major fine-tuning issue for the coupling parameter in order to remain consistent with observations (the impact of which is detailed in \cref{ssec:finetuning}).

In practice, the most stringent bounds on such interactions are expected from the tightly constrained and well observed electromagnetic sector. We thus focus on a coupling of the field $\phi$ to the electromagnetic field curvature tensor of the form 
%In principle, a coupling of the field $\phi$ to the electromagnetic kinetic sector of the form
\begin{equation}
\mathcal{L}_{\rm EM} = -\frac{1}{4}B_F(\phi)F_{\mu \nu}F^{\mu \nu}~,
\end{equation}
as a Lagrangian addition to the full action of \cref{eq:full-action}. Such a term naturally preserves the $U(1)$ gauge symmetry of the theory, and has been adopted in the past in the context of the Bekenstein model \cite{bekensteinOriginal,Olive2002} (we explored such coupled models in detail in \cite{Vacher2022}, using a similar methodology as the present paper). In such a framework, the fine structure constant $\alpha=e^2/4\pi$ is not a constant anymore, and varies with the field as $\alpha(z)/\alpha(z=0) \propto B_F(\phi(z))^{-1}$\,.
%could be considered as acceptable as it preserves the U(1) gauge invariance of the theory. In such a framework, the fine structure constant $\alpha=e^2/4\pi$\footnote{In natural units ($c=\hbar=1$).} -- quantifying the intensity of the electromagnetic interaction -- is not a constant anymore and varies with the field as $\alpha/\alpha_0 \propto B_F(\phi)^{-1}$ (where $\alpha_0\simeq1/137$ is the value of the fine structure constant measured in laboratories on Earth). 
As no significant variation of $\alpha$ can be observed on cosmological scales, it is common to Taylor-expand the coupling at first order as
\begin{equation}
 B_F(\phi) \simeq 1+\zeta\cdot (\phi-\phi_0)~,
\end{equation}
with $\phi_0 = \phi(z=0)$ and $\zeta=\partial_\phi B_F(\phi)|_{\phi=\phi_0}$~, using the sign convention of \cite{Olive2002}. We confirm that since in our runs the field typically does not have large trans-Planckian movement this approximation remains valid. From here we can directly obtain the variation of the fine-sturcture constant as
\begin{equation}\label{eq:variation_alpha}
     \frac{\alpha(z)-\alpha(z=0)}{\alpha(z=0)} \simeq -\zeta \cdot \left[\phi(z)-\phi_0\right]~.
\end{equation}
In our investigation we will simply define $\alpha(z=0)=\alpha_0$\,, where $\alpha_0 \simeq 1/137$ is the value measured in earth-based laboratories \cite{PDG2020}. Furthermore, taking the derivative of \cref{eq:variation_alpha} with respect to $\ln a$, one finds the time variation of the fine structure constant
\begin{equation}\label{eq:alphadrift}
    \frac{1}{\alpha_0}\frac{d \alpha}{d \ln a} = -\zeta \frac{d \phi}{d \ln a}~,
\end{equation}
which can be directly measured in laboratory with high accuracy.

\section{Methodology and data \label{sec:methodology}}

\subsection{Implementation of Quintessence models}
We implement the quintessence model in the Einstein-Boltzmann solver {\sc class}\footnote{\url{https://github.com/lesgourg/class_public}}\cite{Lesgourgues:2011CLASS}. We make use of the existing \texttt{scf} implementation by introducing a new component \texttt{cqf} (coupled quintessence field) with essentially the same equations of motion (which are those of \cref{eq:cqf_kg,eq:cqf_perturbed_kg}). The main differences with respect to the  \texttt{scf} implementation are the different underlying potentials, the initialization of the field, the ability to allow for additional coupling to electromagnetism, and some minor details on the parameter shooting. We list the potentials that we implemented and investigate in \cref{sec:results}. The code is adapting in each case the amplitude of the potential in order to find a solution where the quintessence field makes up a given amount of dark energy ($1-\Omega_m$) in order to achieve a flat universe (as favored from observations and inflationary theories).

The initialization of the field is always performed by providing $\phi_\mathrm{ini}$ and $\dot{\phi}_\mathrm{ini}$ at the initial time at which the {\sc class} background module beings computation. For all investigated cases we explicitly set $\dot{\phi}_\mathrm{ini}=0$ as a 'natural' initial condition. We vary $\phi_\mathrm{ini}$ freely except where explicitly stated otherwise. Investigations with non-zero field speed initial conditions are left for future work. 

The additional coupling to electromagnetism is implemented using \cref{eq:variation_alpha}, where we assume for simplicity a constant coupling $\zeta$. The time-variations of the fine structure constant $\alpha(z)$ are then self-consistently propagated to all parts of the code, such as for example during recombination (building on our previous analysis in \cite{Vacher2022,Vacher2023,Lee2023} and made publicly available in the {\sc class} software), and are used as an output when combining with the data on $\alpha(z)$ or $\dot{\alpha}(z)$ that is described below.

We confront our quintessence models to data by running Monte Carlo Markov chains (MCMC) using the {\sc MontePython}\footnote{\url{https://github.com/brinckmann/montepython_public}} software that is integrated with {\sc class}. Contour plots are made using {\sc getdist}\footnote{\url{https://getdist.readthedocs.io/en/latest/}}. All chains have a Gelman-Rubin convergence criterion of $|R-1| < 0.05$.

\subsection{Data}

The data we employ for our investigation of the cosmological constraints on the Swampland criteria are detailed below:
\begin{itemize}
    \item \textbf{Planck}: The temperature and polarization CMB angular power spectra from the Planck 2018 satellite data release, using the Planck legacy likelihood\footnote{\url{https://pla.esac.esa.int/}}
    \cite{Planck2018V}, as well as the Planck lensing reconstruction \cite{Planck2018VIII}.
    \item \textbf{BAO}: The baryonic acoustic oscillation (BAO) measurements of the angular diameter distance to the sound horizon standard ruler. This combination consists of the old DR12 BAO employed in \cite{Planck2018V}, namely the 6dFGS \cite{Beutler2011} at $z=0.1$, the MGS \cite{Ross2015} at $z=0.15$, and the DR12 LRGs \cite{Alam2017} at $z=0.38,0.51,0.61$.
    \item \textbf{BAO DR16}: The newly released measurements of the BAO from SDSS DR16 \cite{Alam2021}, namely in LRGs (luminous red galaxies) \cite{Gil-Martin2020,Bautista2021} at $z=0.38,0.51,0.7$, in QSOs (quasi-stellar-objects) \cite{Hou2021,Neveux2020} at $z=1.48$, in the Lyman-$\alpha$ forest at $z=2.3$, and their correlation \cite{Bourboux2020}.
    \item \textbf{Pantheon}: The Pantheon supernova sample analysis from \cite{Scolnic2018} containing 1048 supernovae up to $z<2.3$. These standard candles effectively give a measurement of the expansion rate at low redshift.
    \item \textbf{PantheonPLUS}: The updated Pantheon supernova sample analysis from \cite{Scolnic2022} containing 1550 supernovae also up to $z<2.3$.
    \item \textbf{CC}: Cosmic chronometer data from \cite{Moresco2022}, which are measurements of differential ages of galaxies to determine $H(z) \approx -\Delta z / [(1+z) \Delta t]$. We take into account systematic uncertainties outlined in \cite{Moresco2022} such as from the modeling of the star formation history.
    \item \textbf{ACT+SPT}: The ground-based ACT and SPT data from \cite{ACT2020,SPT2023}, which provide the most precise non-Planck CMB anisotropy data to date.
    \item \textbf{QSO} (quasi-stellar-objects) We use spectroscopic measurements of quasars, allowing to recover the value of the fine structure constant $\alpha$ at different redshifts from \cref{eq:variation_alpha}. Our dataset -- detailed in \cite{Martins2017review,Vacher2022} -- contains more than 300 measurements obtained by the VLT-UVES and Keck-HIRES instruments, as well as points obtained from the HARPS, Subaru and ESPRESSO spectrograph, spanning a redshift range of $0.22 \leq z \leq 4.18$ \cite{alphaSubaru,alphaWebb,alphaespresso}.
    \item \textbf{Oklo} is a natural nuclear reactor which was active on Earth several billion years ago. From its abundances, one can obtain a bound of\footnote{This value is obtained from nuclear physics considerations which in turn are very model dependent; specifically, this bound is obtained on the assumption that $\alpha$ is the only dimensionless coupling whose value might not be the standard one. However, as we find later, this is not critical for our study as the final Oklo impact on $\zeta$ is of the same order of magnitude than the QSO data.} $\Delta \alpha/\alpha = (0.005  \pm 0.061)\times 10^{-6}$ on the fine structure constant value of \cref{eq:variation_alpha} at a redshift of $z=0.14$ \cite{Oklo}. 
    \item \textbf{AC} (atomic-clocks) By comparing the transition lines of different atoms in a laboratory, it is possible to obtain the highly restrictive limit on the drift rate of $\alpha$ from \cref{eq:alphadrift} of $\mathrm{d}\ln\alpha/\mathrm{d}\ln a= (2.5\pm3.5)\times10^{-9}$ at $z=0$
    \cite{Filzinger2023}.
\end{itemize}

All the bounds related to the fine structure constant (QSO, Oklo and AC) are implemented as Gaussian prior likelihoods. We also combine the data from Planck, ACT, SPT, CC, BAO DR16 and PantheonPLUS into a data set $\mathcal{A}$ (all). In order to test the impact of new and old data, we also create the dataset $\mathcal{A}_\mathrm{old}$ where we replace BAO DR16 with BAO, and PantheonPLUS with Pantheon.

\section{Results}\label{sec:results}

In this section we present the results of our investigations into the effect of the Swampland criteria \cref{eq:swampland_distance,eq:swampland_desitter} on late-time cosmology and in regard to fine-tuning arguments for the coupling. We start the discussion by presenting constraints for specific models in \cref{ssec:models}, continue in \cref{ssec:finetuning} with a discussion of the extreme fine-tuning required for the coupling of a quintessence field with electromagnetism expected generically, and finish our discussion in \cref{ssec:modelindepndent} with a model-independent analysis of quintessence in the context of Swampland conjectures and a word of caution in regard to this analysis. We use the notation of $\lambda(z)$ and $g(z)$ from \cref{eq:def_lambda}.
\subsection{Model-specific investigation}\label{ssec:models}
\begin{figure}
    \centering
    \includegraphics[width=0.6\textwidth]{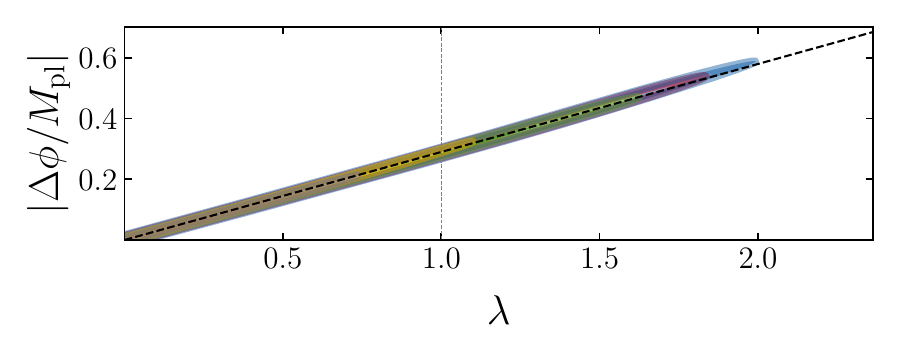}
    \caption{Correlation between the two Swampland criteria in the exponential case}
    \label{fig:cqf_correlation}
\end{figure}
To start our investigation on the coupled quintessence field in the context of the Swampland conjecture, we first revisit some of the commonly considered models. We show the latest constraints on the various models, and note a particularly interesting feature comparing old and new BAO and supernova data.
\subsubsection{Exponential model}\label{sssec:exponential}
Perhaps the most interesting potential is the simple exponential model with
\begin{equation}
    V(\phi) = A \exp(\lambda \phi/\mpl)~.
\end{equation}
Here $\lambda$ is a constant, and we do indeed have $\lambda(z) \equiv \mpl |V'/V| = \lambda$. The exponential case is thus a case where the first Swampland criterion is taken to be constant. This can be argued to be the most conservative choice \cite{Raveri2019} when considering primarily the first Swampland criterion, as any other potential also includes regions with smaller $\lambda(z)$. The second Swampland criterion has $V''/V = \lambda^2 $ implying $g(z) = -\lambda^2 < 0$, which can obviously never be positive, and is thus always violated. It also has an interesting relation to the distance criterion, which is shown in \cref{fig:cqf_correlation} and has been derived analytically in \cite{Raveri2019}. This relation is 
\begin{equation}\label{eq:lambda_eff_distance}
    |\Delta \phi/\mpl| \approx \frac{\lambda}{3} \left[\frac{1}{\sqrt{1-\Omega_m}} \ln\left(\frac{1+\sqrt{1-\Omega_m}}{1-\sqrt{1-\Omega_m}}\right)-2\right] \approx 0.29\lambda~,
\end{equation}
where we used $\Omega_m \approx 0.31$. It is a result of the field being frozen due to Hubble drag during most of the history and only beginning to move when dark energy begins to dominate. Though, since this time-frame is relatively short, the overall displacement of the field is closely related to the slope of the potential during dark energy domination. This mapping immediately allows us to conclude that we primarily have to investigate the first Swampland criterion for this model, since the distance criterion will always be more easily satisfied than the first Swampland criterion, and the second Swampland criterion is by definition violated. As a final note, since we can always absorb a constant offset of $\phi \to \phi + p$ into a redefinition of the field amplitude $A \to A \exp(-\lambda p/\mpl)$ we can fix the initial condition to $\phi_\mathrm{ini} = 0$.
\begin{figure}
    \centering
    \includegraphics[width=\textwidth]{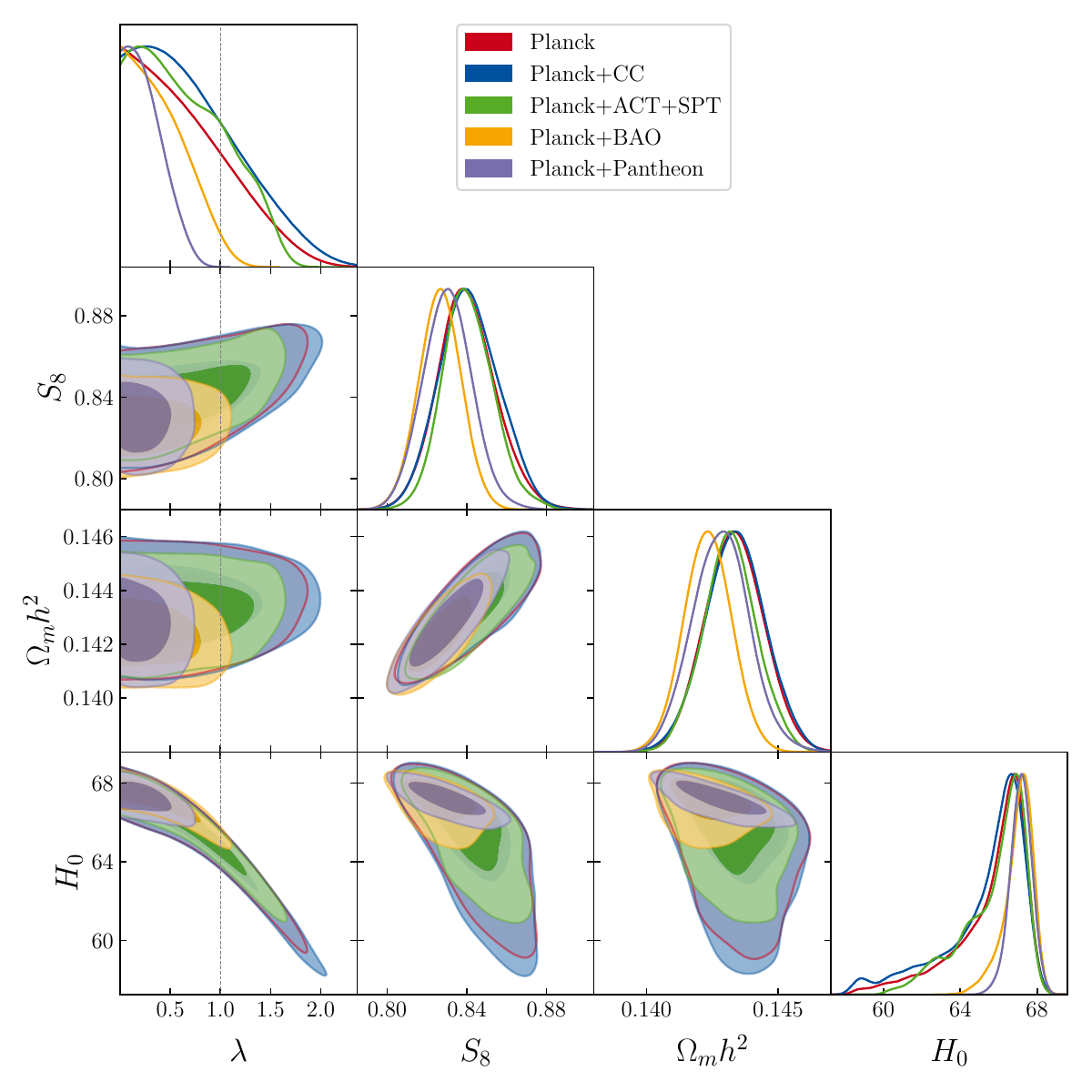}
    \caption{Comparison of the constraints on the exponential model for a variety of data sets.}
    \label{fig:cqf_summary}
\end{figure}
\begin{table}
    \centering
    \begin{tabular}{c|c c c c}
        Data combination & $H_0$ [km/s/Mpc] & $\lambda$ (95\% CL) & $P(\lambda>1)$ in \% & $P(\lambda>0.1)$ in \% \\ \hline
        \rule{0pt}{2.5ex}
        Planck& ${65.6}^{+1.6}_{-1.9}$ & $<1.6$ & 24 & 91\\ \rule{0pt}{2.5ex}
        Planck +CC & ${65.3}^{+1.8}_{-2.2}$ & $<1.7$ & 28 & 93\\ \rule{0pt}{2.5ex}
       Planck +ACT+SPT & ${65.8}^{+1.5}_{-1.7}$ & $<1.4$ & 24 & 91 \\ \rule{0pt}{2.5ex}
       Planck +BAO & $67.02 \pm 0.72$ & $<0.94$ & 2.9 & 85\\ \rule{0pt}{2.5ex}
       Planck +Pantheon & $67.15 \pm 0.56$ & $<0.62$ &$<0.001$ &  79
    \end{tabular}
    \caption{Constraints on $H_0$ and $\lambda$ in the exponential coupled quintessence field model. We also cite the probability to exceed $\lambda>1$.}
    \label{tab:cqf_summary}
\end{table}

We subject the exponential model to a variety of data in \cref{fig:cqf_summary}, showing precisely which cosmological probes give a tight constraint on the model and which do not. We also summarize the most interesting constraints in \cref{tab:cqf_summary}. We observe that while cosmic chronometer data in principle probe the background expansion history, they do not allow for a tight constraint on the field evolution that would be noticeable in a tightened constraint on $\lambda$. Instead, the results are equivalent (or even slightly worse due to finite MCMC convergence). Instead, ACT+SPT data slightly restrict the highest values of $\lambda$. Even more constraining are BAO data which probe the late evolution and the matter fraction $\Omega_m$ tightly, reducing also the probability of $\lambda>1$ with these data to below 5\%. The most constraining data here are the Pantheon supernovae, which allow for a constraint of $\lambda < 0.62$ (95\% CL) when combined with Planck. The probability of reaching $\lambda > 1$ in this model is so low that from the more than 2 million sampled points none have $\lambda>1$. For this reason we do not assign a formal significance\footnote{The MCMC chains based on the Metropolis-Hastings algorithm are not necessarily incredibly well converged at this 4-5$\sigma$ deviation level, and a huge number of samples would be required to make a more quantitative statement.}, but it can be conservatively estimated to be at least smaller than $10^{-5}$. These numbers can be compared to \cite{Raveri2019}, who find less strict bounds both from their CMB data (possibly since they use older Planck 2015 data). The bounds when including supernovae are remarkably similar (they find $\lambda<0.64$ at 95\% CL), but for the conversion into probability they find $1.7\cdot 10^{-4}$ while we find a result smaller than $10^{-5}$. This could be very well caused by the stronger constraint of the newer Planck data on such models in the high-$\lambda$ tail of the distribution.

\begin{figure}
    \centering
    \includegraphics[width=\textwidth]{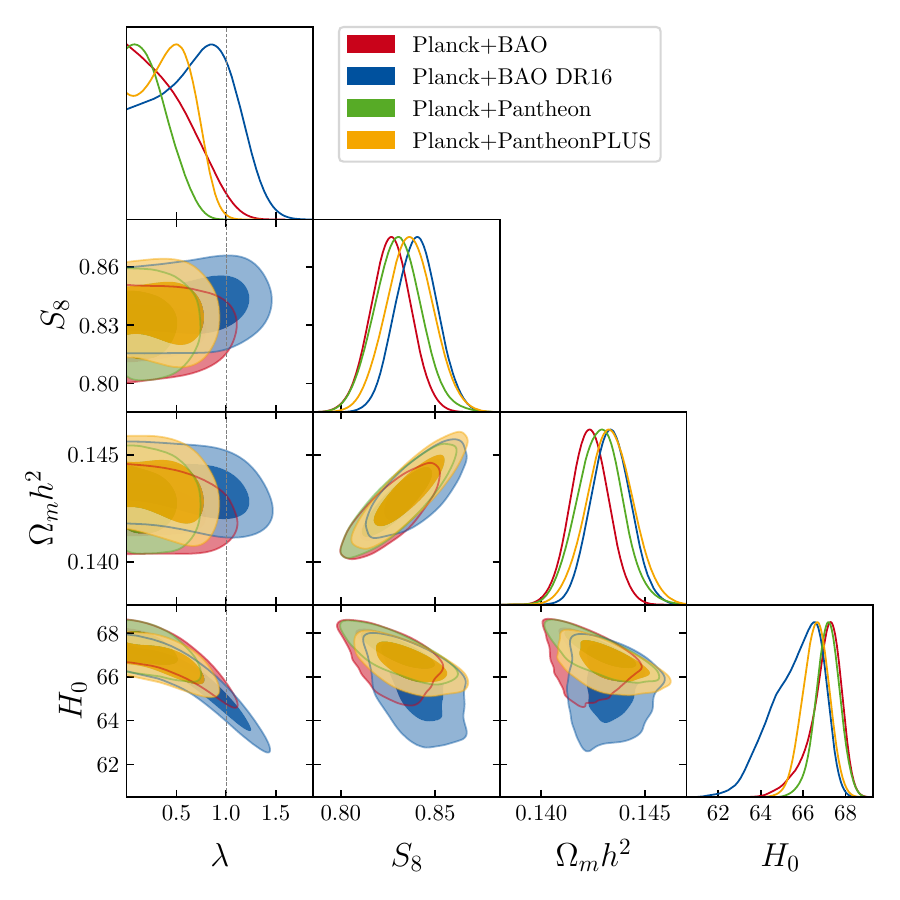}
    \caption{Comparison of the constraints on the exponential model compared between older and newer late-time cosmological data (BAO and SN).}
    \label{fig:cqf_old_new}
\end{figure}
\begin{table}
    \centering
    \begin{tabular}{c|c c c c}
        Data combination & $H_0$ [km/s/Mpc] & $\lambda$ (95\% CL) & $P(\lambda>1)$ in \% & $P(\lambda>0.1)$ in \% \\ \hline
        \rule{0pt}{2.5ex}
        Planck +BAO & $67.02 \pm 0.72$ & $<0.94$ & 2.9 & 85\\ \rule{0pt}{2.5ex}
        Planck +BAO DR16 & $65.7 \pm 1.2$ & $<1.3$ & 23 & 94\\ \rule{0pt}{2.5ex}
        Planck +Pantheon & $67.15 \pm 0.56$ & $<0.62$ & $<0.001$ & 79\\ \rule{0pt}{2.5ex}
        Planck +PantheonPLUS & $66.59 \pm 0.62$ & $<0.80$ & 0.24 & 89
    \end{tabular}
    \caption{Same as \cref{tab:cqf_summary} for old and new BAO and Pantheon data.}
    \label{tab:cqf_old_new}
\end{table}

To check how much this is impacted by using newer BAO and SN data, we show in \cref{fig:cqf_old_new} a comparison of the older and newer data. Surprisingly, the newer data has looser bounds on the $\lambda$ parameter. While at first very puzzling, this does make sense. The newer PantheonPLUS data in particular prefers relatively high values of $\Omega_m$ compared to the older data, resulting in a lower abundance of the dark energy component, which is then less restricted in $\lambda$ due to its lower abundance. We summarize these constraints also in \cref{tab:cqf_old_new}. 

\begin{figure}
    \centering
    \includegraphics[width=\textwidth]{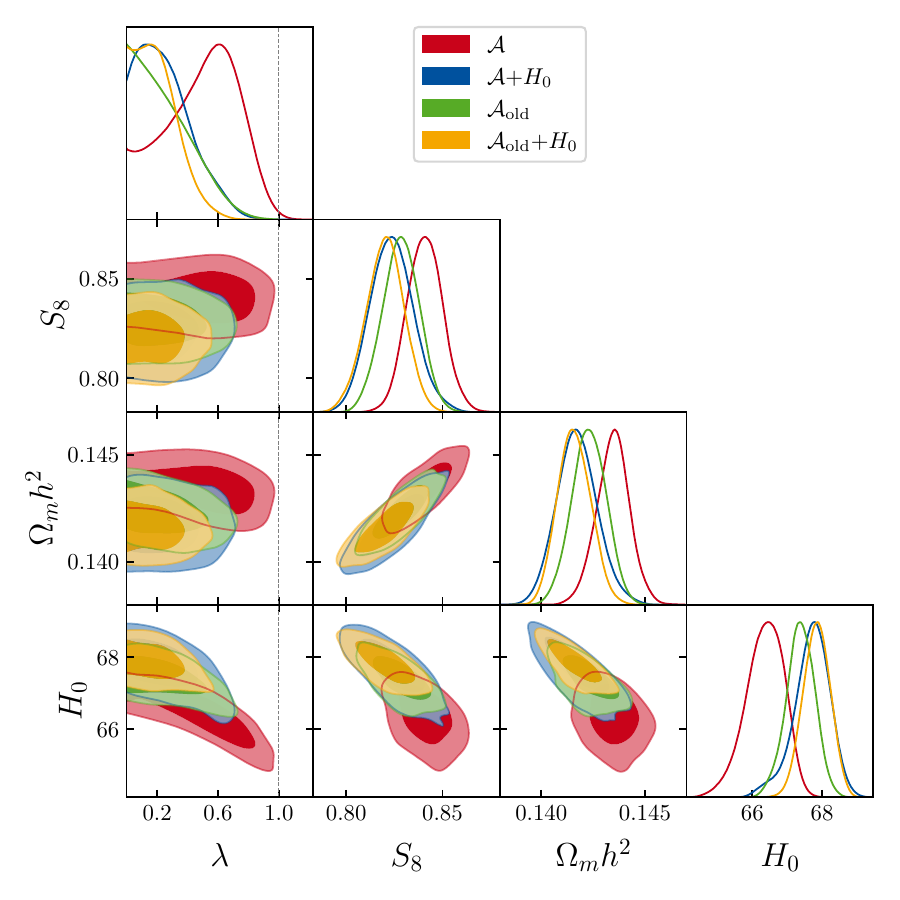}
    \caption{Comparison of the constraints on the exponential model for a variety of all data.}
    \label{fig:cqf_all}
\end{figure}
\begin{table}
    \centering
    \begin{tabular}{c|c c c c}
        Data combination & $H_0$ [km/s/Mpc] & $\lambda$ (95\% CL) & $P(\lambda>1)$ in \%  & $P(\lambda>0.1)$ in \% \\ \hline
        \rule{0pt}{2.5ex}
        $\mathcal{A}$& ${66.36}^{+0.56}_{-0.54}$ & $<0.85$ & 0.33 & 0.93\\ \rule{0pt}{2.5ex}
        $\mathcal{A}+H_0$ & $67.70 \pm 0.50$ & $<0.60$ & $<0.001$ & 0.80\\ \rule{0pt}{2.5ex}
        $\mathcal{A}_\mathrm{old}$ & $67.38 \pm 0.40$ & $<0.60$ & $<0.001$ &0.77 \\ \rule{0pt}{2.5ex}
        $\mathcal{A}_\mathrm{old}+H_0$  & $67.84 \pm 0.39$ & $<0.46$ & $<0.001$ & 0.74
    \end{tabular}
    \caption{Same as \cref{tab:cqf_summary} for combinations involving the $\mathcal{A}$=Planck, BAO, Pantheon, CC, ACT, and SPT data combination, and also for added $H_0$ priors.}
    \label{tab:cqf_all}
\end{table}

\pagebreak[10]
When applying our dataset $\mathcal{A}$ (involving Planck, BAO, Pantheon, CC, ACT, and SPT data), we find that the results are fairly similar, see \cref{fig:cqf_all} (indicating that BAO and supernovae data are dominating the constraining power). Here too the old data is more restrictive and has tighter constraints. We observe that in almost all cases $\lambda$ and $H_0$ are anti-correlated, since a lower $H_0$ generally corresponds to a higher matter fraction $\Omega_m$ due to the tight constraint on $\Omega_m h^3$ from the Planck satellite measurement of the sound horizon angular scale. In turn, the higher matter fraction corresponds to a lower abundance of the coupled quintessence field ($\Omega_\phi \approx 1-\Omega_m$) allowing for larger values of $\lambda$ to be viable. As such, we also show how the constraints change when including a prior on $H_0$ from \cite{Riess2022}. In this case, we find that the constraints on $\lambda$ are a lot tighter, which given the above discussion is not entirely surprising. However, given the high tension between $H_0$ derived from CMB data and from the local-distance-ladder, this result should be taken with a grain of salt. We summarize the results in \cref{tab:cqf_all}.

\subsubsection{Cosine potential}

\begin{figure}
    \centering
    \includegraphics[width=0.9\textwidth]{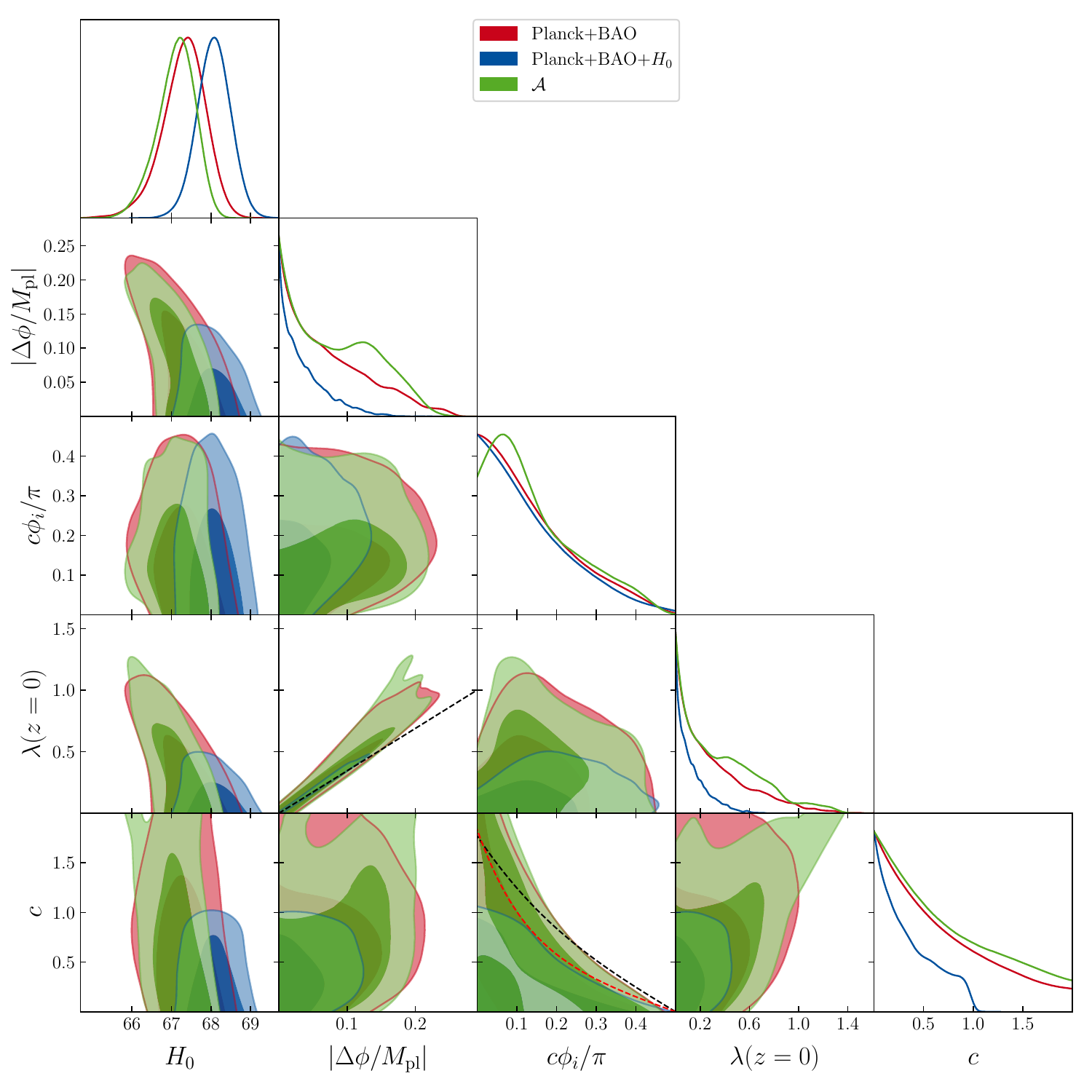}
    \caption{Constraints on the model parameters for the cosine potential model. The black dashed line in the $\lambda(z=0)$\,-\,$|\Delta \phi/\mpl|$ panel is the relation $|\Delta \phi/\mpl| = 0.29 \lambda(z=0)$. The black and red dashed line in the $c$\,-\,$c \phi_i/\pi$ panel represents \cref{eq:lambda_cosine_relation} for $\lambda_\mathrm{eff}=1$ and $\lambda_\mathrm{eff}=0.5$, respectively.}
    \label{fig:cqf_cosine}
\end{figure}
\begin{figure}
    \centering
    \includegraphics[width=0.45\textwidth]{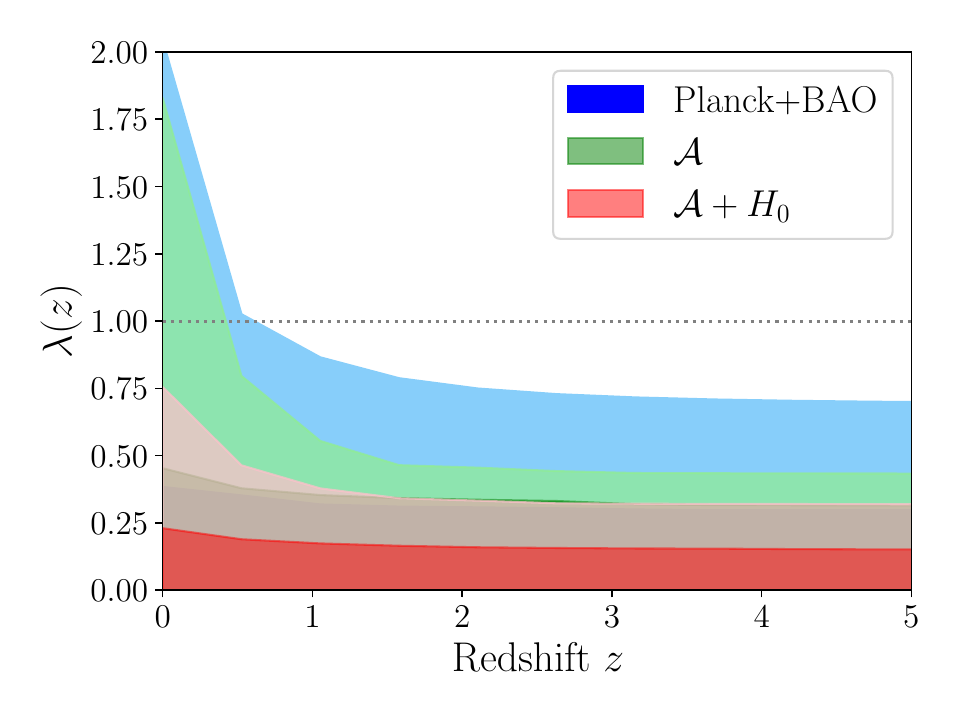}
    \includegraphics[width=0.45\textwidth]{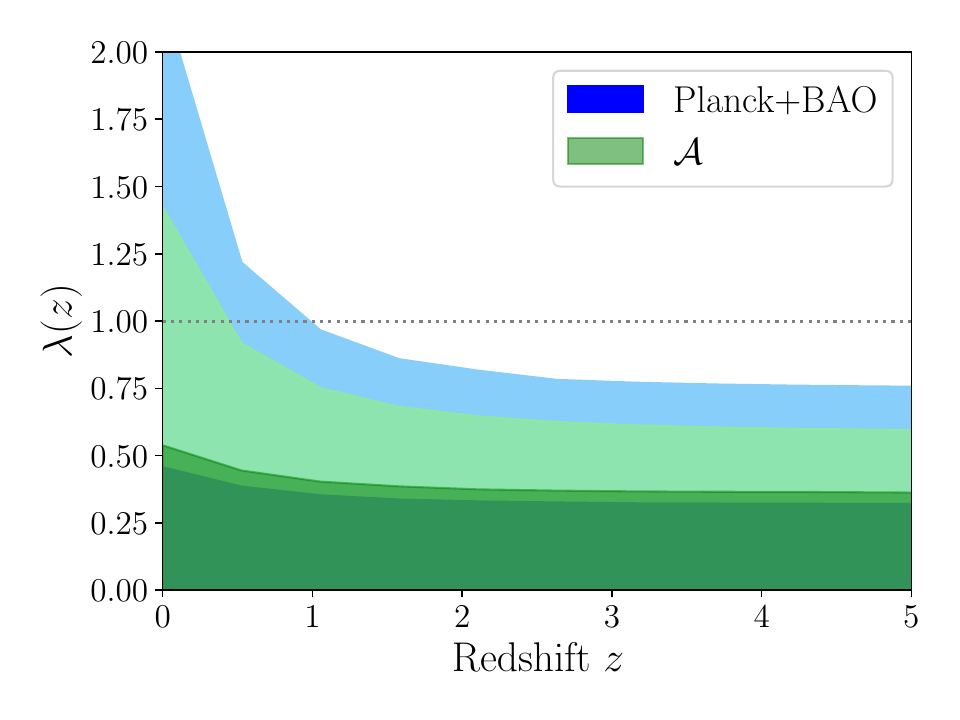}
    \caption{Constraints (68\% and 95\% upper limits) on $\lambda(z)$ in the cosine and axionic cosine potential. The blue contours are from Planck+BAO data, while the green contours involve the dataset $\mathcal{A}$. The red contours add also a $H_0$ prior from \cite{Riess2022}.}
    \label{fig:reconstr_cos_axcos}
\end{figure}
Another potential with a constant Swampland criterion is given by a cosine, where the second Swampland criterion is set to some constant ($g(z) = c^2$), resulting in a potential of the type
\begin{equation}
    V(\phi) = A \cos(c \phi/\mpl)~.
\end{equation}
This case is also sensitive to the initial value of $\phi$, since there is no scaling symmetry here, and as such we also vary the initial value. The resulting constraints are displayed in \cref{fig:cqf_cosine}. 

As pointed out by \cite{Raveri2019}, since this model is initially frozen by Hubble drag and only begins moving during the dark energy dominated epoch (thawing model), the same tight correlation between the (relative) slope $\lambda(z)$ and the overall displacement can be observed as in the exponential case. Indeed, this is what we show in \cref{fig:cqf_cosine}, using $\lambda(z=0)$ as a proxy for the late-time slope. However, the relation is not perfect if the overall displacement $|\Delta \phi/\mpl|$ becomes large. Indeed, one can derive using $|\Delta \phi/\mpl| \simeq 0.29 \lambda_\mathrm{eff}$ with some effective slope $\lambda_\mathrm{eff}$ (note that \cite{Raveri2019} use a weighted average of $\lambda(z)$ by $\rho_\phi/\rho_\mathrm{crit}$) what the relationship between the initial field $\phi_i$ and the cosine curvature $c$ should be. Indeed, by definition $\lambda(z=0) = c \tan(c \phi_i/\mpl + c \Delta \phi/\mpl)$. Approximating $\lambda(z=0)$ by $\lambda_\mathrm{eff}$ and using \cref{eq:lambda_eff_distance} one can find 
\begin{equation}\label{eq:lambda_cosine_relation}
    \phi_i \approx -0.29 \lambda_\mathrm{eff} + \arctan(\lambda_\mathrm{eff}/c)/c~,
\end{equation}
which is the curve we show in \cref{fig:cqf_cosine} for $\lambda_\mathrm{eff}=1$ (black) and $\lambda_\mathrm{eff}=0.5$ (red). As we can observe, the true constraints are deviating from this line, being especially more permissive for large values of $c$. Indeed, this is also noticeable in the deviation of the scaling of $|\Delta \phi/\mpl|$ compared to $0.29 \lambda(z=0)$. For large values of $c$ the field traverses more different potential slopes for a given distance that it travels, breaking the tight correlations -- these are not visible in \cite{Raveri2019} due to only showing the tightest dataset, which in our case would be similar to the blue contours, which do nicely follow the relation (compare for example the blue contours in the $c$\,-\,$c \phi_i/\pi$ panel to the red dashed line, and the tight correlation in the $|\Delta \phi/\mpl|$\,-\,$\lambda(z=0)$ panel). In this more general case we see the deviations from the simple scaling laws derived in \cite{Raveri2019} especially for large field displacements (large $|\Delta \phi/\mpl|$) and fast oscillations (large $c$). Since adding $H_0$ data to an otherwise incompatible (at $\sim 5\sigma$) collection of datasets is a rather questionable choice and can artificially drive constraints, we prefer to show the constraints mostly without the $H_0$ prior (shown in blue), and thus these deviations become important.

To summarize, the cosine potential case is very much related to the exponential case (with constant slope) especially for small $c$ or small $\Delta \phi$ when the field cannot experience much more than the local slope during the cosmologically short dark energy domination period. The tight relations of the exponential case hold here too, except when $c$ or $\Delta \phi$ become large. A reconstruction of the $\lambda(z)$ can be found for completeness in \cref{fig:reconstr_cos_axcos}, which shows that there is a tight bound on $\lambda(z)$ during most of the history, but the $\lambda(z)$ is allowed to significantly grow towards $z \to 0$. Due to this later growth, the initial bound on $\lambda$ is also tighter compared to the exponential case.

\subsubsection{Axionic Cosine potential}

\begin{figure}
    \centering
    \includegraphics[width=0.7\textwidth]{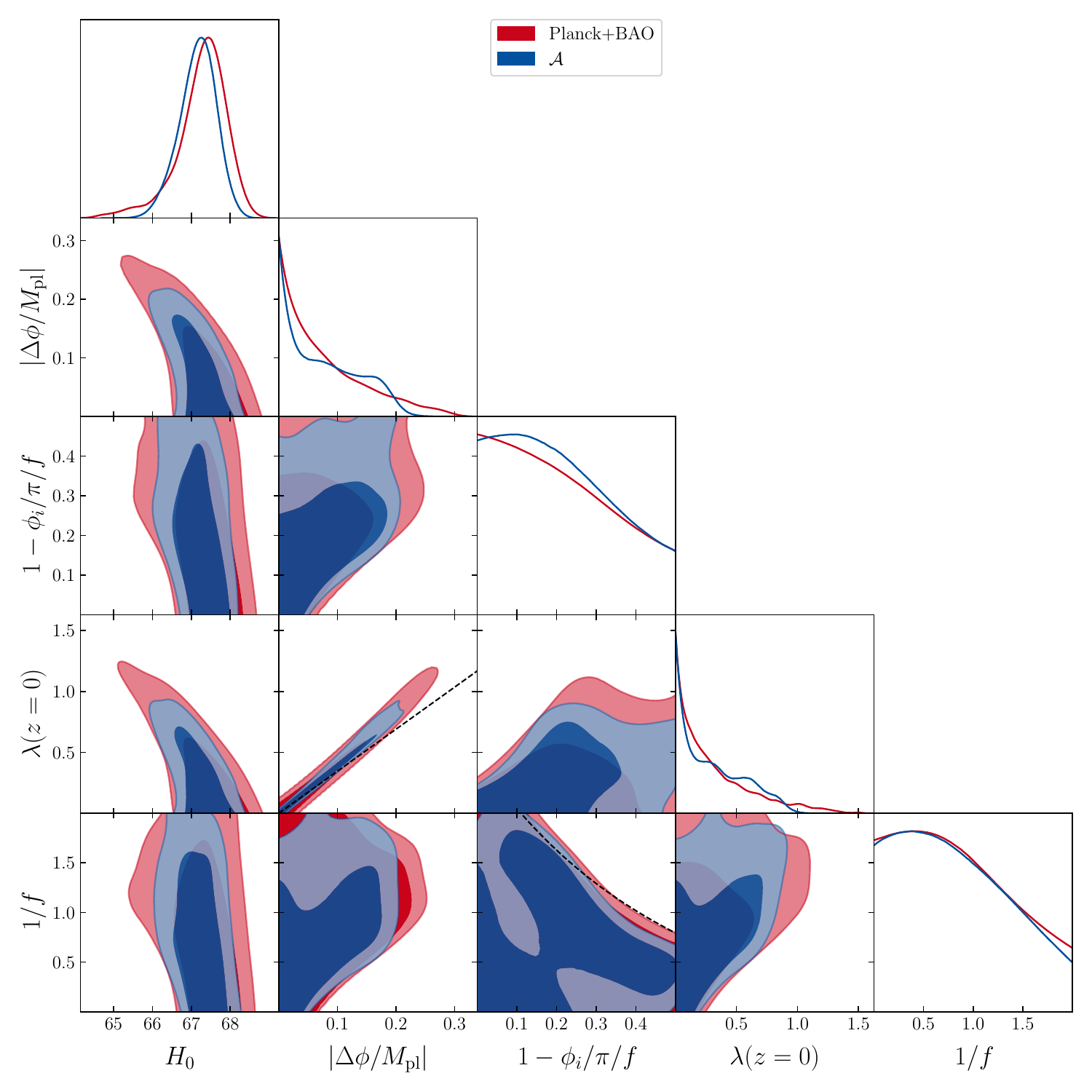}
    \caption{Same as \cref{fig:cqf_cosine} but for the axionic cosine potential of \cref{eq:axcos_potential} instead. }
    \label{fig:cqf_axcos}
\end{figure}
In order to check also for more natural potentials, we investigate a toy model of the effective axionic potential (e.g. adopted in \cite{Kamionkowski2014,Poulin2019}, which is the instanton approximation\footnote{Looking at \cite[Eq. (10)]{diCortona2016}, since the up and down quarks obey $4 m_u m_d/(m_u+m_d)^2 \approx 0.6$, the approximation is decent as long as $m$ is allowed to be varied freely (giving a 3\% mean square error approximation for $m \approx 1.3\,m_a$). As the resulting constraints are so similar to the normal cosine case, we do not expect significant changes from a more accurate modeling.} \cite{diCortona2016}) as
\begin{equation}\label{eq:axcos_potential}
    V(\phi) = m^2 f^2 \left[1-\cos(\phi/f) \right]~,
\end{equation}
where $m$ is an amplitude parameter and $f$ is a frequency parameter. This model behaves very similarly to the cosine potential one (featuring also $g(z) \approx \mathrm{const}$) since a field starting close to $\phi = f\pi - \phi_i$ will undergo a very similar evolution as one starting at $\phi=\phi_i$ in a cosine potential with $c = \mpl/f$. A corresponding computation of the relation between $\phi_i$ and the effective slope yields
\begin{equation}\label{eq:lambda_axcos_relation}
    \pi - \phi_i \approx -0.29 \lambda_\mathrm{eff} + \frac{1}{c}\arctan_2\left(\left.\frac{-2 c \lambda_\mathrm{eff}}{c^2+\lambda_\mathrm{eff}^2}\right|\frac{c^2-\lambda_\mathrm{eff}^2}{c^2+\lambda_\mathrm{eff}^2}\right)
\end{equation}
where $\arctan_2(x|y) = \{ \arctan(y/x) \text{~for~$x>0$}, \pi+\arctan(y/x) \text{~for~$x<0$}\}$ with $c=1/f$, which is shown to be a good fit in \cref{fig:cqf_axcos}. Given the high degree of  similarity between this case and the ordinary cosine, we can draw essentially the same conclusions in this case. The only modification is a slight adjustment of where the $c$\,-\,$c \phi_i/\pi$ bounds occur, due to the difference between \cref{eq:lambda_cosine_relation,eq:lambda_axcos_relation}. The reconstructed constraints on $\lambda(z)$ are shown in \cref{fig:reconstr_cos_axcos} and are essentially identical to those of the cosine potential. It should be noted that in this case $g(z)\neq c^2=1/f^2$. Indeed, a reconstruction of $g(z)$ gives the approximate bound $g(z) < 1.6$ (95\% CL) in this case, though it is not clear whether this is impacted by the chosen priors on $1/f$ (and as such we do not show the reconstruction).

\subsubsection{Inverse exponential}
\begin{figure}
    \centering
    \includegraphics[width=0.8\textwidth]{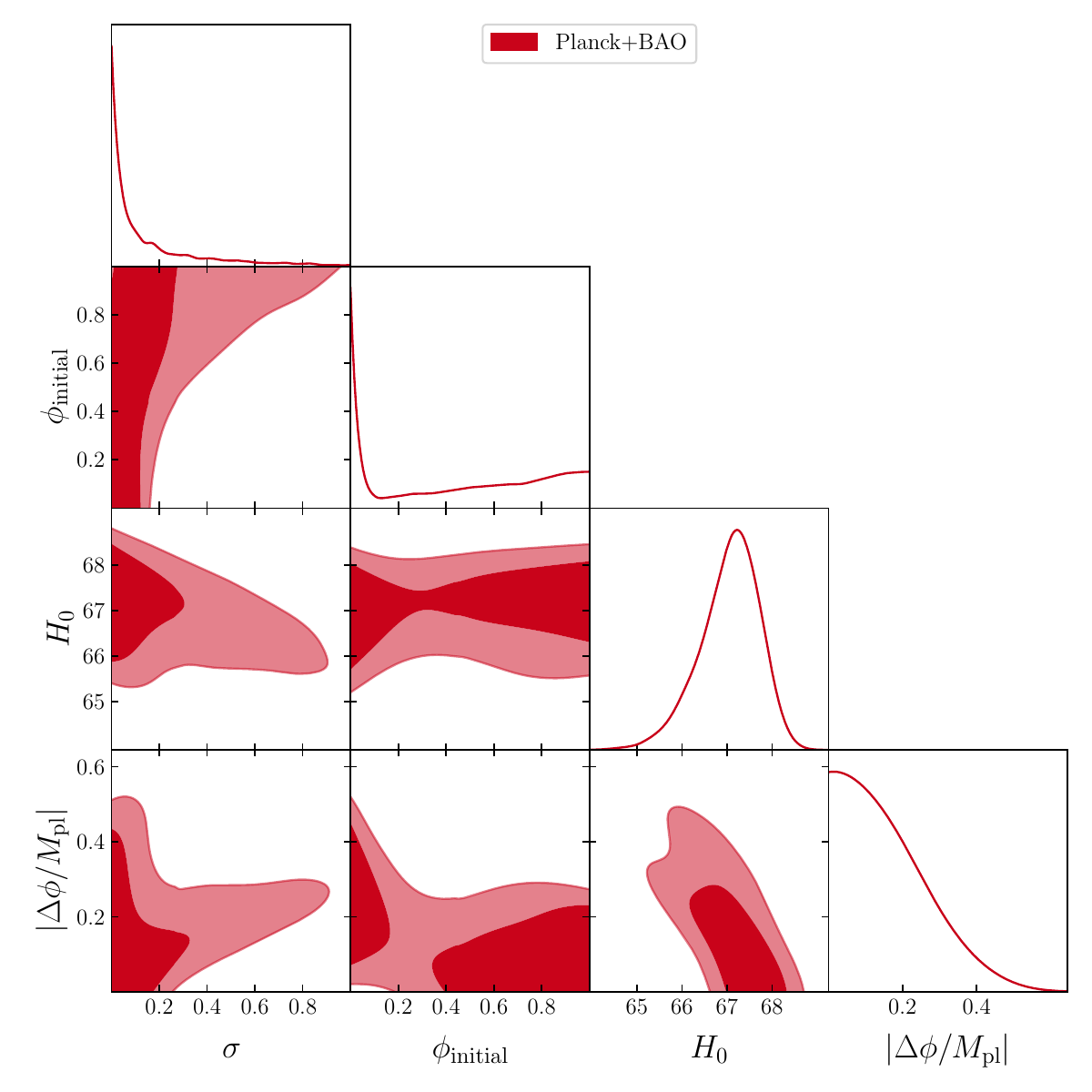}
    \caption{Constraints for the inverse exponential case.}
    \label{fig:inv_exp_summary}
\end{figure}
\begin{figure}
    \centering
    \includegraphics[width=0.45\textwidth]{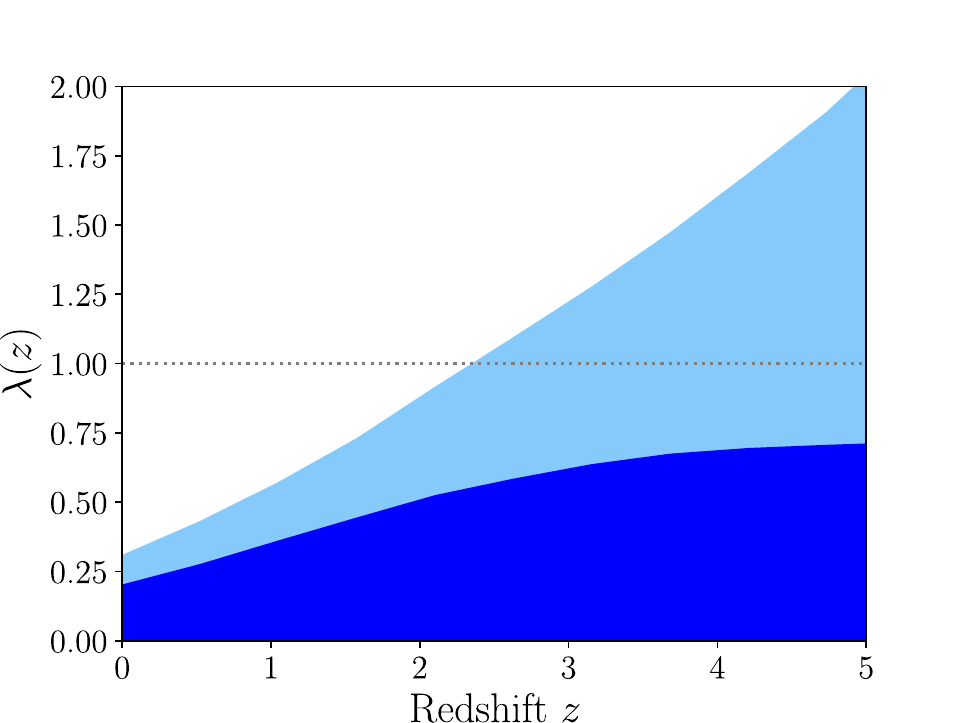}
    \includegraphics[width=0.45\textwidth]{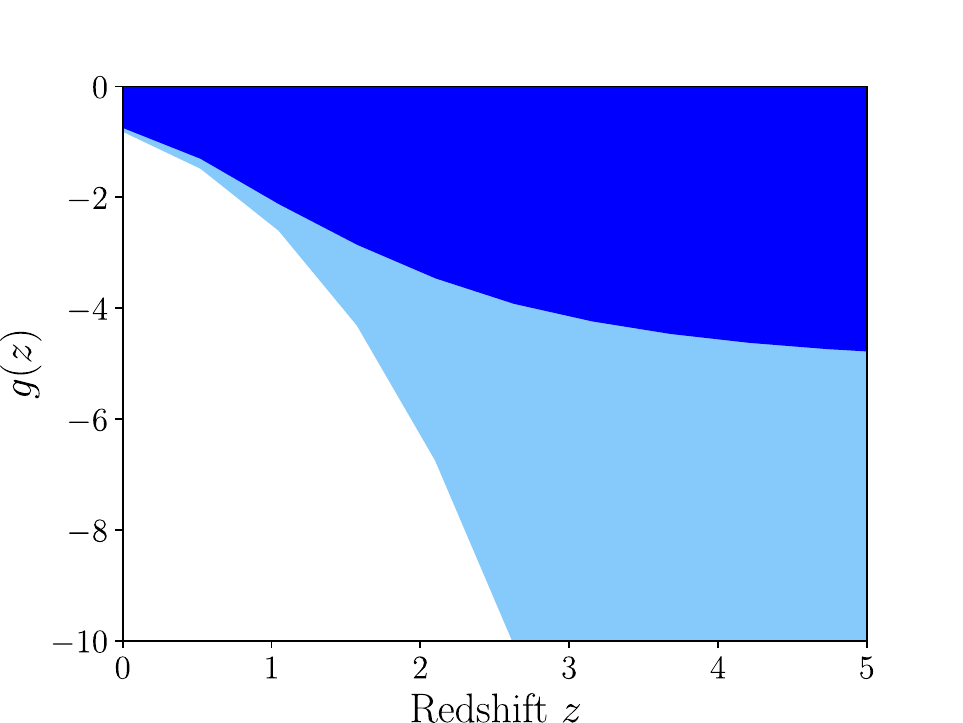}
    \caption{Reconstruction of the two Swampland criteria for the inverse exponential potential. For the first criterion we show the 1 and 2 sigma upper bound, for the second criterion we show the 1 and two sigma lower bound (since it is negative by definition, and thus the second criterion is always violated).}
    \label{fig:reconstr_inv_exp}
\end{figure}

This potential is defined as
\begin{equation}
    V(\phi) = A \exp(\sigma \mpl/\phi)
\end{equation}
and features the possibility of the dark energy thawing into an attractor solution. We show the constraints for the parameters of this potential in \cref{fig:inv_exp_summary}. There is a curious bi-modality for the initial field displacement. In the low initial field displacement, the field is quickly kicked from its initial value and rolls until it overshoots the attractor solution and stops at a high displacement, before eventually rejoining the attractor solution, hence showing large displacements. On the other hand, high initial field values do not accelerate very much and are thus found with small final displacements. For the same reason higher values of $\sigma$ are allowed for these fields. Overall, the correlation with the Hubble constant is mild, and the field offset is not dramatic. We also show a reconstruction of the slow roll parameters in \cref{fig:reconstr_inv_exp}. We note that as in the exponential case all curves feature $g(z) < 0$ and thus always violate the second Swampland criterion. Interestingly this model is also much more restrictive for the first Swampland criterion, $\lambda(z)$, though the constraint is very quickly deteriorating for high redshift. This is in line with the statement that the exponential case is the most conservative case for low redshift where the dark energy dominates, but also shows that the added flexibility in the inverse exponential case allows it to explore higher $\lambda(z)$ at high redshift inaccessible to many other models.

\subsubsection{Double exponential}

\begin{figure}
    \centering
    \includegraphics[width=0.8\textwidth]{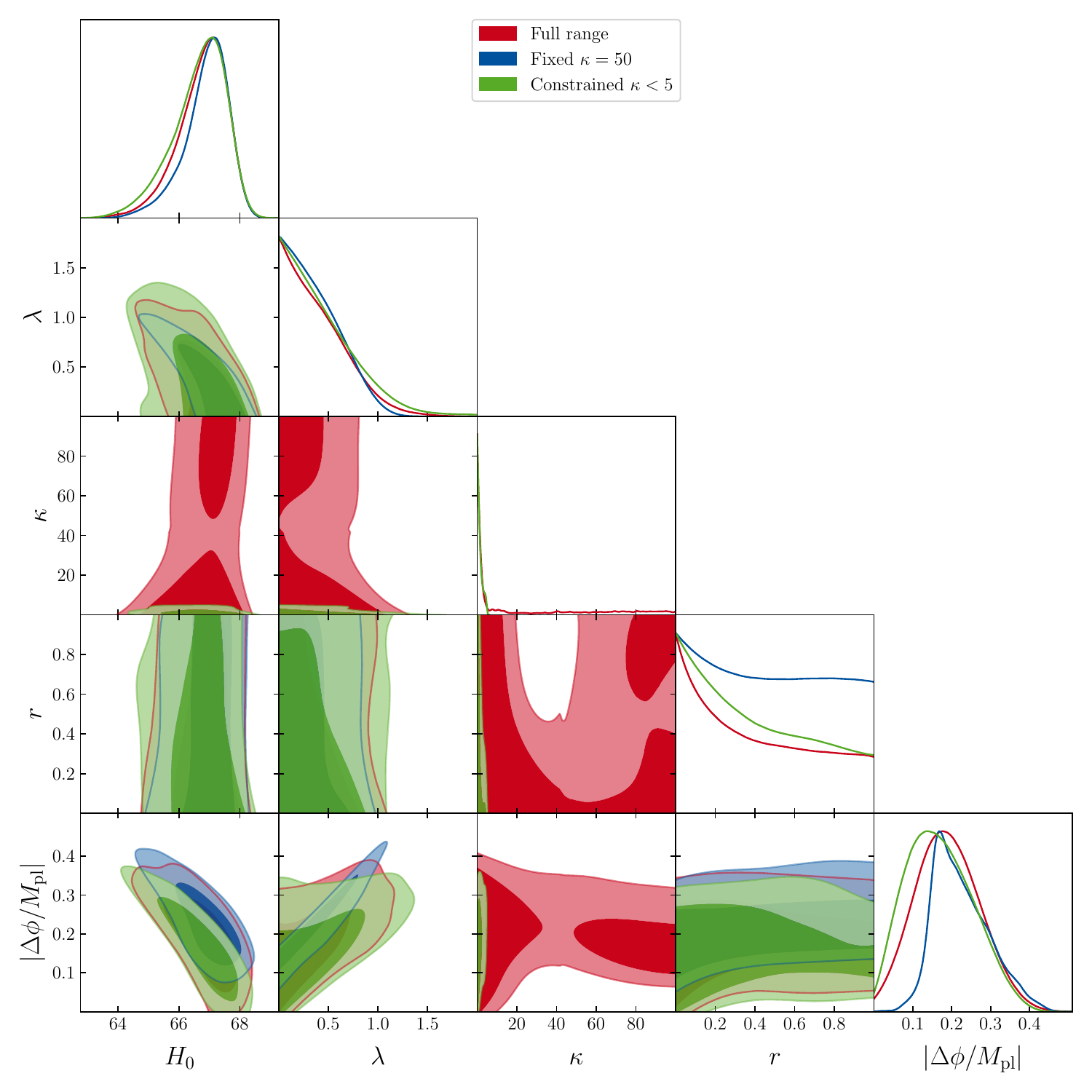}
    \caption{Constraints for the double exponential case.}
    \label{fig:double_exp_summary}
\end{figure}
\begin{figure}
    \centering
    \includegraphics[width=0.8\textwidth]{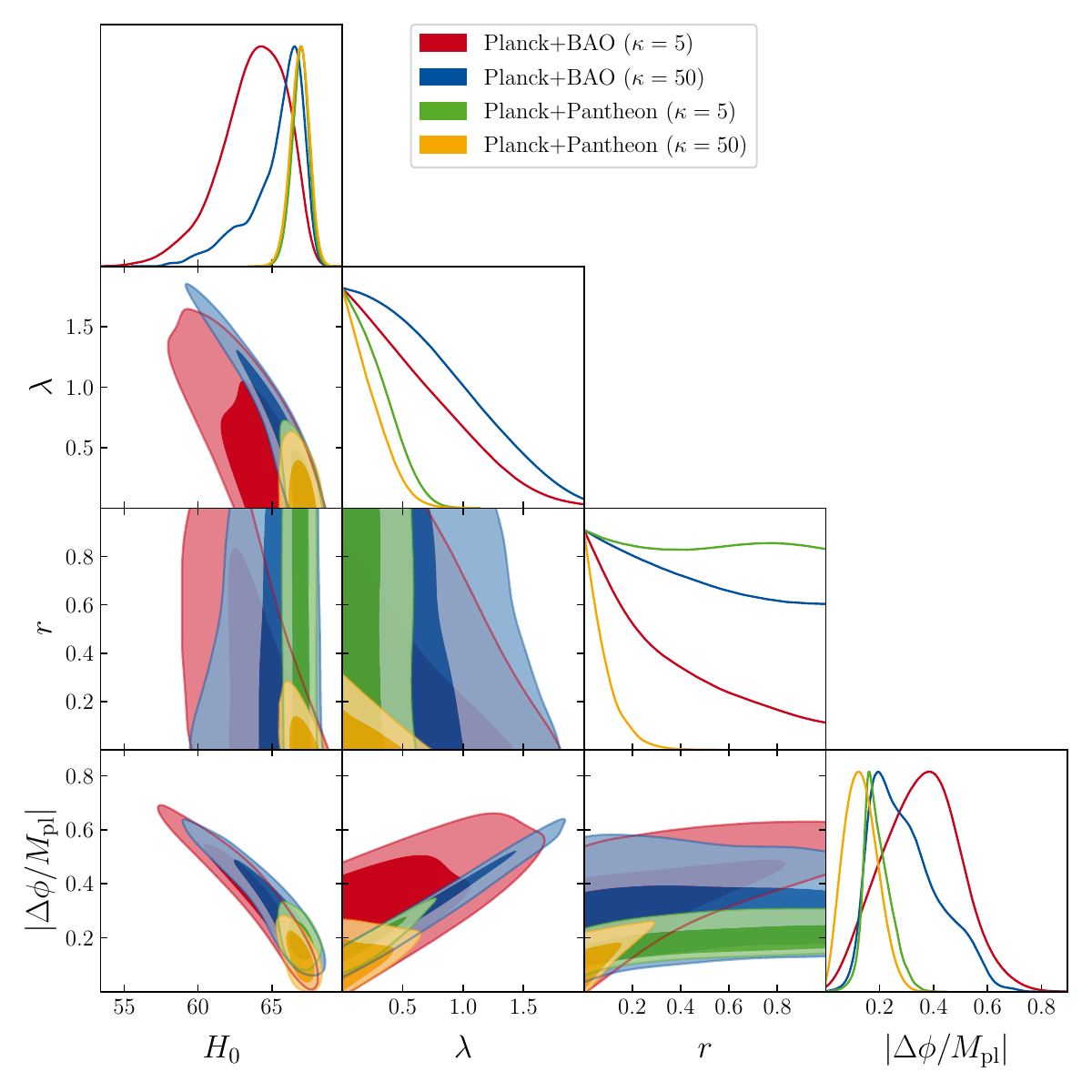}
    \caption{Constraints for the double exponential case, with either a fixed low or fixed high value of $\kappa$.}
    \label{fig:double_exp_summary_variations}
\end{figure}

\begin{figure}
    \centering
    \includegraphics[width=0.6\textwidth]{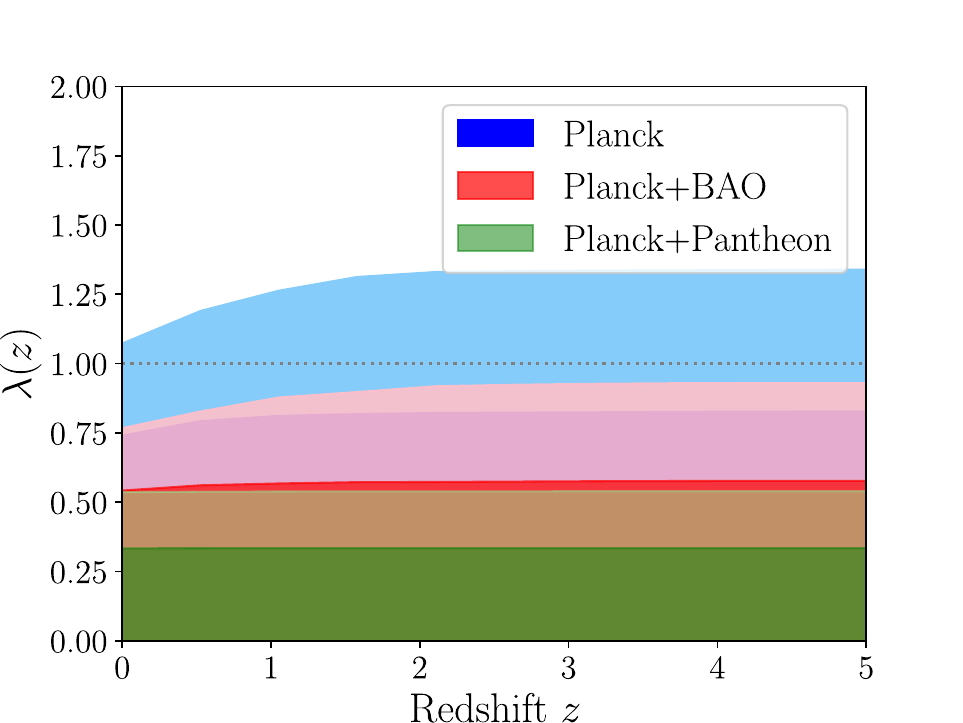}
    \caption{Low $\kappa$ reconstruction of the double exponential case, clearly showing that the the effect of the second exponential is at most minute, and the constraints on $\lambda$ in these cases can be seen already as sufficient.}
    \label{fig:reconstr_double_exp}
\end{figure}

The double exponential potential is defined as 
\begin{equation}
    V(\phi) = A \left[\exp(\lambda \phi/\mpl) + r \exp(\kappa \phi/\mpl) \right]
\end{equation}
where $r$ is the ratio of the two exponential amplitudes and $\kappa$ is the steepness of the second exponential. Here too a constant offset of the field $\phi \to \phi+p$ can be reabsorbed by  simultaneously redefining $A \to A \exp(\lambda p/\mpl)$ and $r \to r \exp((\kappa-\lambda) p/\mpl)$, so we do not vary $\phi_i$\,.

We show constraints for this model in \cref{fig:double_exp_summary,fig:double_exp_summary_variations}. In a preliminary exploratory run we allow all parameters to vary in large margins, with $r \in [0,1]$ and $\kappa \in [0,100]$. However, the resulting parameter space is highly multi-modal. This is because the regime for large $\kappa$ features one exponential that is very quickly becoming irrelevant for small displacements, leaving the other exponential to carry most of the cosmological impact. Indeed, the constraints on $\lambda$ in this case are almost identical to those of the single exponential case of \cref{sssec:exponential}. Instead, the limit of small $\kappa$ is also interesting where the second exponential can act in the same way as the first one, which allows the two exponentials to conspire to create more complicated potential shapes. Nevertheless we still find very competitive constraints for $\lambda$ even in this case. This is because we do not allow the second potential to dominate ( $r \gg 1$ is excluded by our priors ), and thus the first potential will always significantly contribute to the field history, hence resulting in a tight constraint on $\lambda$ throughout.  It is interesting to note that around $\kappa=50$ the tightest constraints on $r$ are seen, and it is not quite clear why $r$ is so constrained in this range. It seems that this is simply a sweet-spot where the second potential has a big impact on the overall expansion history. We separate the cases $\kappa<5$ and $\kappa=50$ as examples for the possible evolution of the double exponential potential, and show the constraints on these cases separately. It is interesting to note that due to the dominance of the first exponential term (since $r \in [0,1]$) the reconstructed Swampland constraint histories are all essentially flat, as can be appreciated in \cref{fig:reconstr_double_exp} for a reconstruction of $\lambda(z)$ in the double exponential case with a prior on $\kappa<5$. We essentially have $\lambda(z) \approx \lambda = \mathrm{const}$ since the other exponential doesn't strongly contribute. This is particularly true once tighter constraints are imposed from BAO or Pantheon data.

\subsection{Fine-tuning arguments}\label{ssec:finetuning}

New scalar degrees of freedom have no reason a priori to be decoupled from all other fields, such as the fields underlying the standard model of particle physics. In particular, the well measured electromagnetic interaction is expected to couple with the quintessence fields if there is not some assumed fundamental reason for the coupling to be precisely zero. Given the constraints outlined below, this leaves us with two alternative problems. Either we have to construct a theoretical argument to prevent such interactions, or we have to face a great amount of fine-tuning of the interaction rate being many orders of magnitude less than what could be naturally expected. For example, one might invoke a global symmetry to suppress these couplings, but this is not a viable solution, bearing in mind that quantum gravity effects do not respect global symmetries and that there are no (unbroken) global symmetries in string theory \cite{Banks:1988yz}. In a nutshell, by predicting that dark energy should be dynamical (and not a cosmological constant), the Swampland criteria also predict a varying $\alpha(z)$.

\begin{figure}
    \centering
    \includegraphics[width=0.75\textwidth]{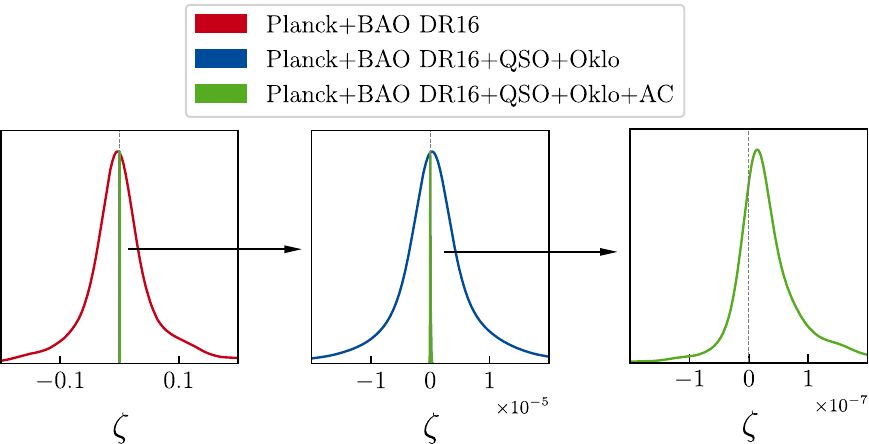}
    \includegraphics[width=0.85\textwidth]{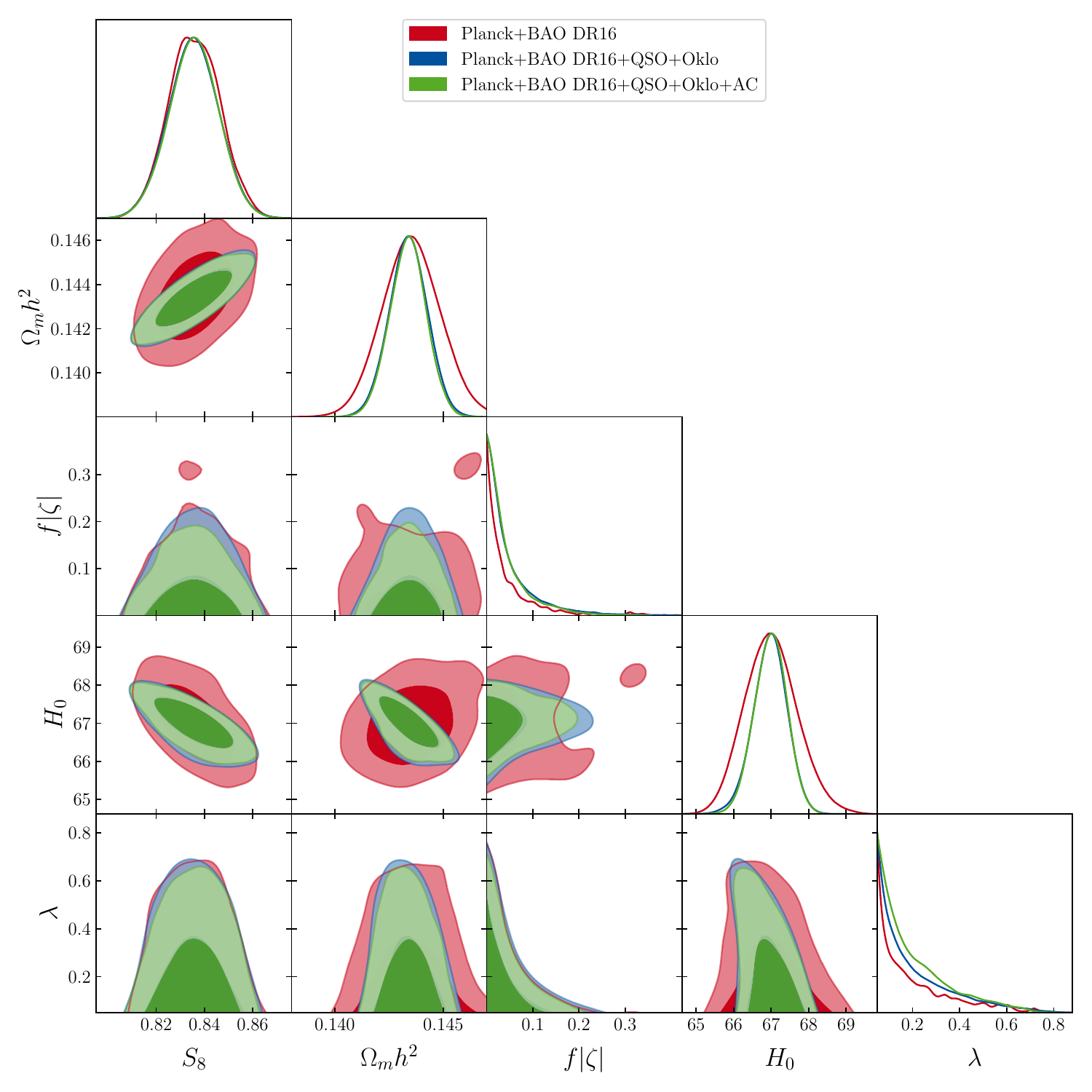}
    \caption{Constraints on $\zeta$ for the coupled and interacting coupled quintessence field. Top: Constraints on $\zeta$ alone, with a zoom-in for the $\alpha(z)$ constraints. Bottom: Full triangle plot with relevant parameters, where we transform $\zeta$ to $f |\zeta|$ with $f=1$ for Planck+BAO DR16, $f=10^4$ when including also QSO+Oklo data, and $f=10^6$ when also including AC data.}
    \label{fig:cqf_interacting}
\end{figure}

The constraints on the interacting quintessence field are displayed in \cref{fig:cqf_interacting}, and summarized in \cref{tab:cqf_interacting}. While the shapes of the degeneracies are very similar with and without the fine structure data, the constraint on the coupling strengthens by multiple orders of magnitude from the observations of $\alpha(z)$. Indeed, couplings of order unity are excluded by these constraints at many thousands of $\sigma$ level significance. As such, the necessary variation of $\alpha(z)$ introduced by any natural coupling would be immediately excluded by astrophysical and laboratory experiments.
\begin{table}[]
    \centering
    \begin{tabular}{c|c}
        Data combination & Constraint on $|\zeta|$ (95\% upper CL) \\ \hline \rule{0pt}{3ex}
        Planck+BAO & 0.19 \\
        Planck+BAO+QSO+Oklo & $1.8 \cdot 10^{-5}$ \\
        Planck+BAO+QSO+Oklo+AC & $1.5 \cdot 10^{-7}$
    \end{tabular}
    \caption{Constraints on the coupling parameter $\zeta$ from the various combinations of data employed within this section.}
    \label{tab:cqf_interacting}
\end{table}

The presence of a space-time variation of the fine-structure constant should also induce a violation of the universality of free-fall (for a review see \cite{Damour2009,Uzan2011}). As such, by constraining the differential acceleration of two nearby geodesics, $\eta$, it is possible to bound the possible values for $\zeta$ \cite{Carroll1998,Dvali2002,Chiba2002}. Typically, in our case, one expects $\eta \sim O(10^{-4}-10^{-3})\zeta^2$, such that the latest MICROSCOPE bound $\eta = (-1.5\pm 2.7) \cdot 10^{-15}$ \cite{Touboul2022} would be less constraining than atomic clocks (as also put forward in \cite{Vacher2022} for the so-called BSBM model, and in \cite{Archidiacono:2022iuu} for an ultra-light scalar).\footnote{Still, the sharp MICROSCOPE bound can play a significant role in some specific varying-$\alpha$ models, further tightening the possible values for the coupling of the field with the rest of the standard model. For a review see \cite{Berge2023}.} However, the non-observation of a differential acceleration puts another independent constraint on any coupling.

In string theory, for which the Swampland conjectures are expected to be maximally motivated, such fields coupled to electromagnetism and leading to a variation of the fine structure constant are expected and unavoidable. This is the case of the moduli fields arising from compactification and more critically from the so-called dilaton field, always present along the graviton as a vibrational mode of the closed strings and setting the value of the string coupling. Several routes have been proposed to face the fine tuning issues set by measurements of the fine structure constant and tests of the universality of free fall. For example, some \enquote{screening} mechanism could be invoked, explaining how the effects of the field could remain unnoticed in a dense environment such as the Earth (see e.g., \cite{terHaar2021}). Moreover,  fields such as the dilaton could gain a significant enough mass (through e.g. supersymmetry breaking) such that their effects would be exponentially quenched with distance. \cite{Damour1994} proposed that some attractor behavior in the matter-dilaton system could lead to the decoupling between the dilaton and the other fields of the standard model, explaining the small values of the couplings today. However, even in this situation, data drastically constrain the possible parameter space of the model, excluding most of its expected and natural values \cite{Martins2019,Vacher2023}. On the other hand, one must also explain how to stabilize the extra compact dimensions and the associated moduli fields, in order to severely restrain their dynamics. 

As such, almost any natural order coupling of the quintessence with electromagnetic fields expected from most unified theories has to face either the large hurdle of constructing a theoretical reason to effectively forbid such a coupling or has to face problems of extreme fine-tuning.

\subsection{A note on model-independent reconstructions} \label{ssec:modelindepndent}

In \cite{Elizalde2019,Yang2020,Arjona2021,Elizalde2021,Park2021,Heisenberg2021} the authors put constraints on the coupled quintessence field in a model-independent way. It is important to stress that model-independence in this context does not mean independence from assumptions about the cosmological evolution. In particular, assumptions about the precise functional behavior of $H(z)$ in the quintessence model are replaced with assumptions about the smoothness of $H(z)$. However, the derived results have to be interpreted in the context of these underlying assumptions.

To outline why these assumptions are so crucial, we perform a reconstruction of expansion rate data from supernovae and cosmic chronometers below, and use this analysis to point out a few of the pitfalls of model-independent reconstructions. The first of these critical assumptions/pitfalls is the data that is used for the reconstruction. It is well known that cosmic chronometers allow for measurements of the Hubble parameter $H(z)$. However, as outlined in \cite{Moresco2020,Moresco2022} a great deal of care is to be taken, as the commonly used $H(z)$ data can be subject to biases if the systematic uncertainties (such as from the stellar population synthesis model) are not properly taken into account. For this section only, we use the data of \cite{Moresco2022} (Tab 1, which is the same as that of \cite{Gomez-Valent2018}) without any further uncertainties or correlations. 

There exist little to no other cosmological data that can currently compliment these tight measurements directly. Instead, the model independent reconstructions typically fall back to the use of indirect determinations of the Hubble parameter, such as through supernovae or BAO. However, neither of these probe the Hubble expansion history directly. Without the calibration of their absolute luminosity $M_B$\,, supernovae only determine the Hubble parameter up to a multiplicative factor. Effectively, they constrain the expansion rate $E(z) = H(z)/H_0$\,. In order to use these data then for constraints on Swampland parameters, one has to either adopt a given value of $H_0$ or of $M_B$ (which is directly equivalent to adopting a value of $H_0$\,, since it can be converted using the same supernovae and the distance ladder approach). Similar issues arise for the use of BAO data, which measure either the angular extent of the BAO feature $\Delta \theta \propto r_s/D_A(z)$ or its redshift extent $\Delta z \propto H(z) r_s$\,. Typically only the latter (the parallel) measurement is used for these reconstructions. Even in this case, the measurement is not directly of $H(z)$, as crucially the sound horizon at drag $r_s$ serves as an overall multiplicative factor. Here too some fiducial value has to be adopted in order to convert the data into $H(z)$ data. This point of requiring a fiducial model to convert $E(z)$ or $H(z) r_s$ data into $H(z)$ data is typically neglected by many reconstructions, but should be kept in mind when interpreting the results. 

\begin{figure}
    \centering
    \includegraphics[width=0.45\textwidth]{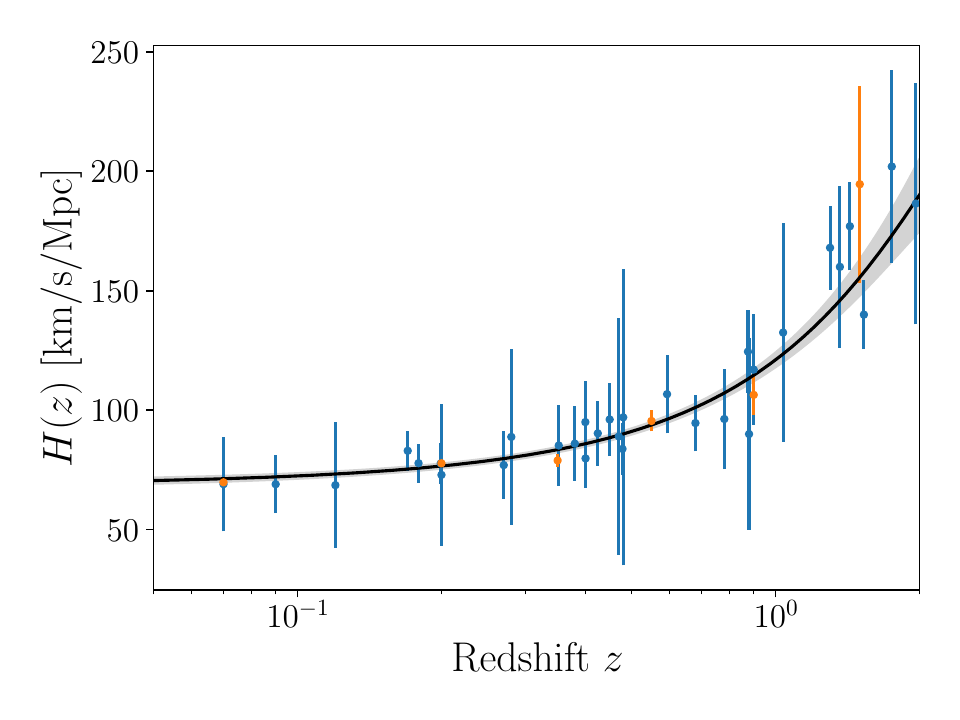}
    \includegraphics[width=0.45\textwidth]{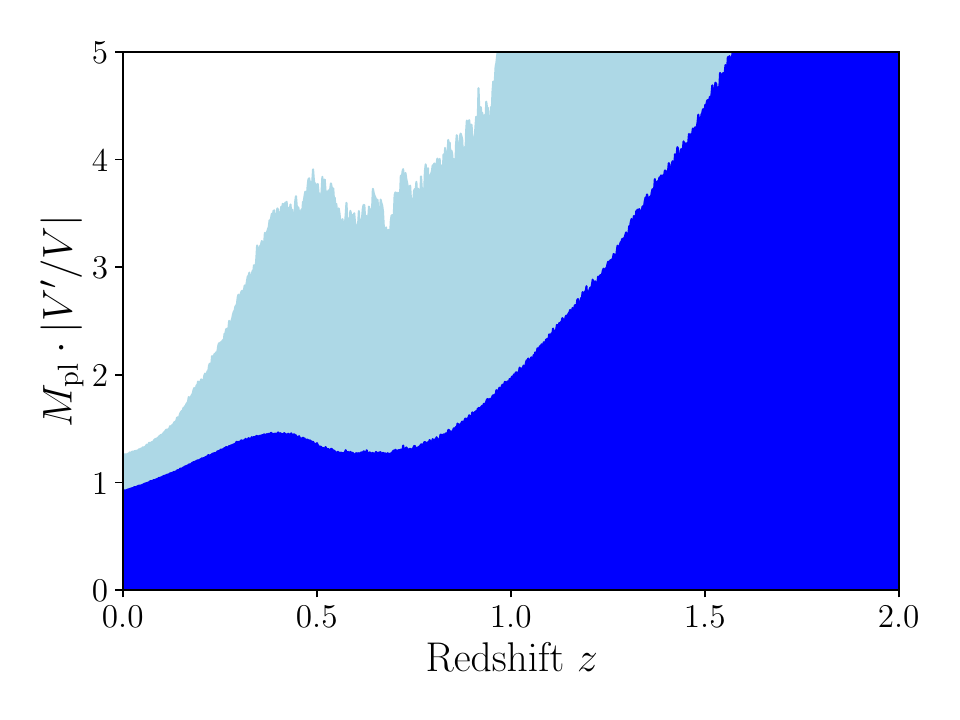}
    \caption{The reconstruction of the first Swampland criterion, using the fiducial assumptions ($H_0=70$km/s/Mpc, $\Omega_m=0.3$). Left: The data used for the reconstruction, as well as the fitted Gaussian Process with a square exponential kernel (mean=black line, uncertainty=grey area). Right: The one-sided upper bounds (68\%, 95\%) from the quantiles of the histories, with the cuts applied on a $z$-dependent basis (see text). Note the similarity to the results of \cite[Fig.~1]{Yang2020} and \cite[Fig.~3]{Elizalde2019}.}
    \label{fig:reconstr_model_indep}
\end{figure}

The data we are going to use for the reconstruction in this work are going to be the $H(z)$ and $E(z)$ data from \cite{Gomez-Valent2018} (tables 1 and 2). We also checked that the addition of BAO data from eBOSS DR16 (LRG, QSO, Lya) \cite{Alam2021,Gil-Martin2020,Bautista2021,Hou2021,Neveux2020} does not significantly change any of the conclusions, and in order not to use any early-universe calibration ($r_s$) we explicitly leave out the BAO data. The data are shown in \cref{fig:reconstr_model_indep}. The first step is then to build an emulator of this $H(z)$ data with error bars. This can either be done using a Gaussian process (this work, \cite{Elizalde2019,Yang2020}), a cosmographic expansion (\cite{Arjona2021,Banerjee2021,Heisenberg2021}), using genetic algorithms (\cite{Arjona2021}), or using other emulation techniques not adopted in these works (such as neural networks). Each of these methods uses explicitly or implicitly an assumption about the smoothness of the interpolating function. For example, cosmographic expansions and genetic algorithms typically assume infinitely differentiable basis functions. For the Gaussian Process regression, the smoothness assumption typically depends on the choice of underlying kernel function, such as the squared exponential kernel (this work), or a given order Matern kernel. This becomes important when considering that these assumptions about smoothness are fundamental when evaluating the derivative of the emulation. Indeed, the Matern-$(p+1/2)$ kernel for the Gaussian process reconstruction only allows derivatives up to order $p$. Explicitly, the Matern-$3/2$ reconstruction would typically be subject to infinite second derivatives, and thus be entirely unable to make a statement about the Swampland conjecture. It is likely for this reason that \cite{Elizalde2019} chose $p=4$ when considering a Matern kernel (which is then already somewhat close to a squared exponential covariance function, which is the limit of the Matern kernel for $p \to \infty$). A reconstruction of the Swampland criteria without assuming a given level of smoothness is thus fundamentally impossible. In the fit with a Gaussian Process, this assumption is perhaps the most explicit, with control over the length scale as a hyperparameter that can be adapted in order to maximize the data likelihood, but in the other cases this assumption of length scales is `baked into' the underlying basis functions of emulation. 

Finally, in order to reconstruct $V(\phi)$ from $H(z)$ and its derivatives, one needs to explicitly subtract the non-quintessence contributions to the expansion rate. This is typically done by explicitly assuming a given value of the matter density $\Omega_m$ and declaring that apart from the quintessence at the redshifts of interest only the standard dust-like matter should be present, i.e. assuming explicitly \mbox{$H^2 = \mpl^{-2} \rho_\phi/3 + H_0^2 \Omega_m (1+z)^3$}. This, for example, explicitly neglects the impact of massive neutrinos. This is admittedly negligible for current neutrino mass bounds, but it does show that there is no complete `model-independence' here. Even further, many of the analyses simply adopt a given value of $\Omega_m$ without further discussion. As we show below, the choice of $\Omega_m$ does have a non-negligible impact on the final constraints.

Even after this subtraction has been performed, there can still be problems. In particular, since now $\rho_\phi = 3\mpl^2 \left[H(z)^2 - H_0^2 \Omega_m (1+z)^3\right]$, there is in principle no guarantee that $\rho_\phi > 0$ for all possible emulations of $H(z)$. This means that either explicitly or implicitly part of the reconstruction space is discarded. This problem becomes even more acute when computing the kinetic term $\dot{\phi}^2 = \rho_\phi + P_\phi$ (see \cref{eq:field_density_pressure}), which is required to convert time-derivatives of the potential into derivatives with respect to the field. This kinetic term involves $\mathrm{d}\rho_\phi/\mathrm{d}z$ (or equivalently $H'(z)$) since the conservation equation implies \mbox{$\dot{\rho_\phi} = - 3 H (\rho_\phi+P_\phi) = -3 H \dot{\phi}^2$}. As such, not only is $\rho_\phi > 0$ required but also $\mathrm{d}\rho_\phi/\mathrm{d}z >0$. As such, for a very general emulation one naturally finds regions where these conditions are violated. In order to perform a viable analysis, either the redshift range of interest or the space of reconstructed models has to be further cut. These cuts often impose an additional prior on the corresponding model-independent bounds that is rarely explicitly discussed. If the reconstruction of $V'/V$, for example, is not compatible at the $1\sigma$ level with zero, this could be due to the data preferring $V'/V > 0$, or it could be a result of the various assumptions and cuts discussed above.

\begin{figure}
    \centering
    \includegraphics[width=0.45\textwidth]{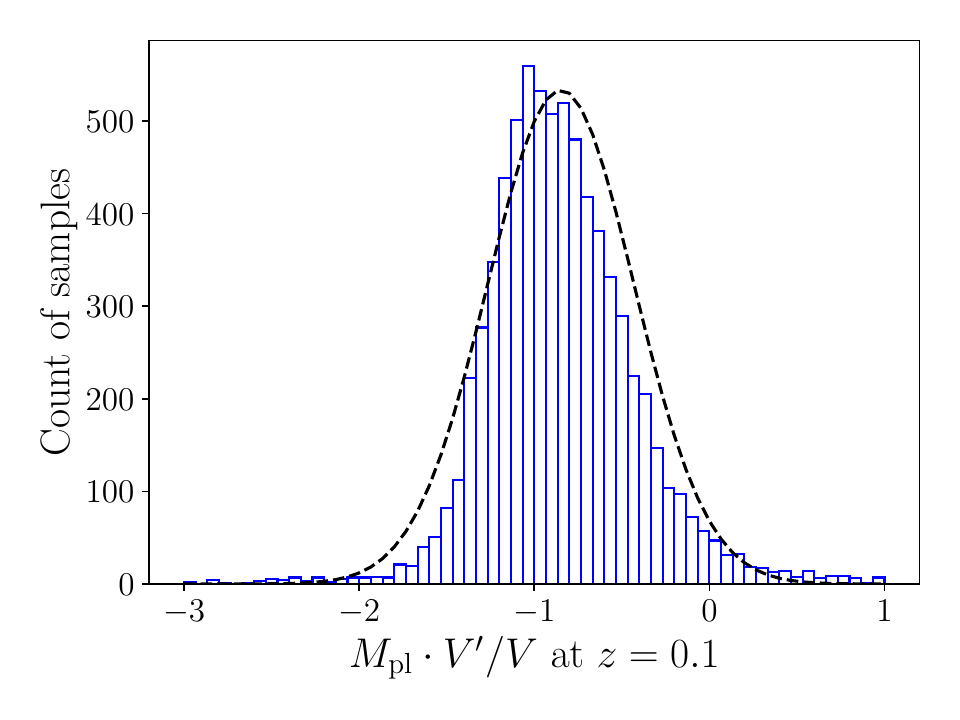}
    \includegraphics[width=0.45\textwidth]{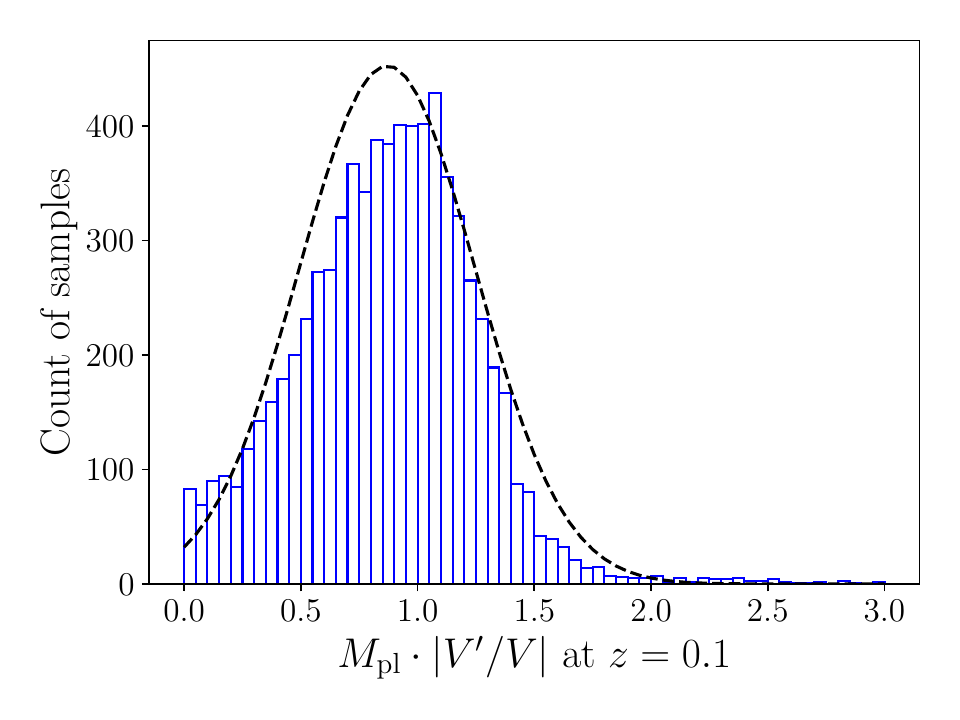}
    \caption{Distribution of samples at $z=0.1$ for the reconstruction of $V'/V$ and $|V'/V|$, showing that both are rather non-Gaussian distributions, especially after taking the absolute value. In blue are the histograms of the individual samples, in black are fitted Gaussian curves. We already eliminated the impact of outliers on the Gaussian fit by restricting to $V'/V \in [-3,1]$ and $|V'/V| \in [0,3]$.}
    \label{fig:reconstr_nongaussian}
\end{figure}

\begin{figure}
    \centering
    \includegraphics[width=0.45\textwidth]{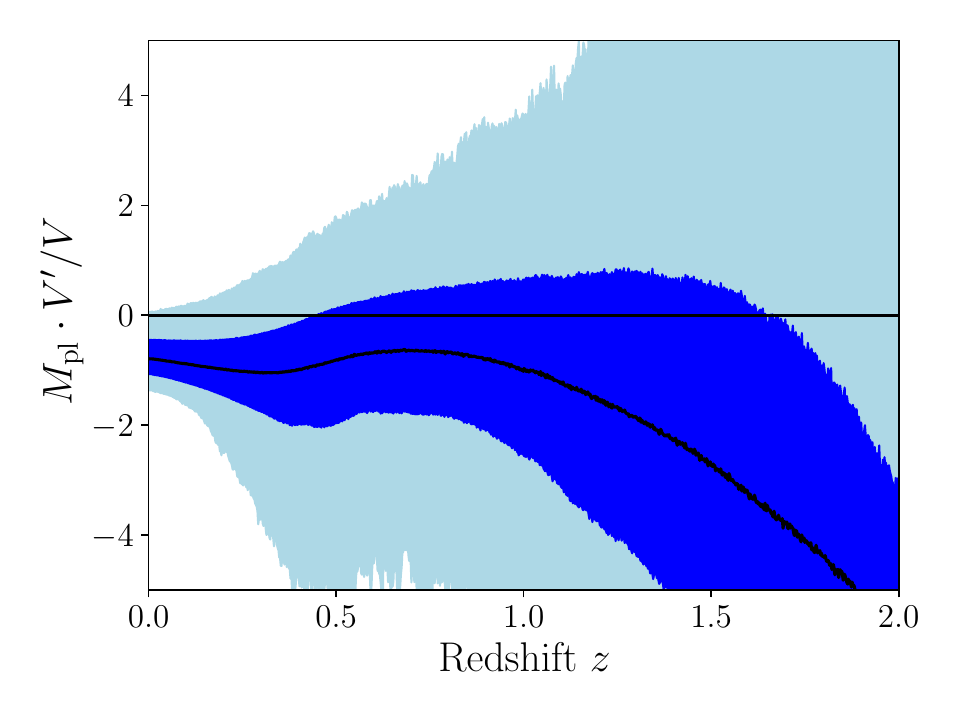}
    \includegraphics[width=0.45\textwidth]{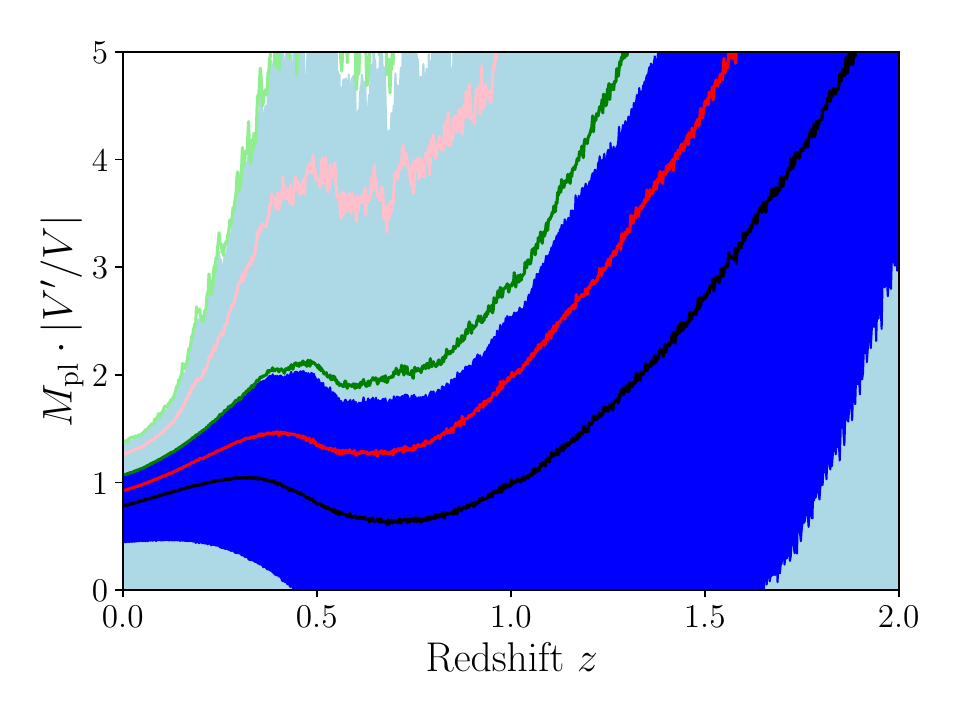}
    \caption{Comparison of techniques for deriving upper/lower quantiles on the reconstruction given a set of samples. Left: $15.87\%-84.13\%$ quantile in dark blue, $2.275\%-97.725\%$ quantile in light blue, and median (50\% quantile) in black. Right: Different possible limits in comparison. The light-blue and dark blue contours as well as the black line are the absolute values of those from the left panel. The red and pink lines are the $68.3\%$ and $95.4\%$ quantiles of the absolute value samples, while the dark green and light green lines are the $84.13\%$ and $97.725\%$ quantiles of the absolute value samples for $|V'/V|$.}
    \label{fig:reconstr_quantiles}
\end{figure}

In our analysis, we sample\footnote{The results from the explicit sampling are the same as those obtained from using mean and covariance from the analytical derivatives of the Gaussian processes (when properly including also their covariance), but it gives us more flexibility about imposing cuts of physicality of $\rho_\phi>0$ and $\mathrm{d}\rho_\phi/\mathrm{d}z >0$.} a total of 10\,000 histories from the Gaussian process that is fit to $H(z)$ and $E(z)$ data described above (the latter including the covariance), using 1000 redshift samples between 0 and 2. We impose the physicality conditions $\rho_\phi>0$ and $\mathrm{d}\rho_\phi/\mathrm{d}z >0$ discussed above for each redshift independently, which in our case is a dramatic cut of around 70\% of the total samples for all histories. Note that the final result does depend on how exactly such a cut is performed, as we explicitly show below.

\begin{figure}
    \centering
    \includegraphics[width=0.45\textwidth]{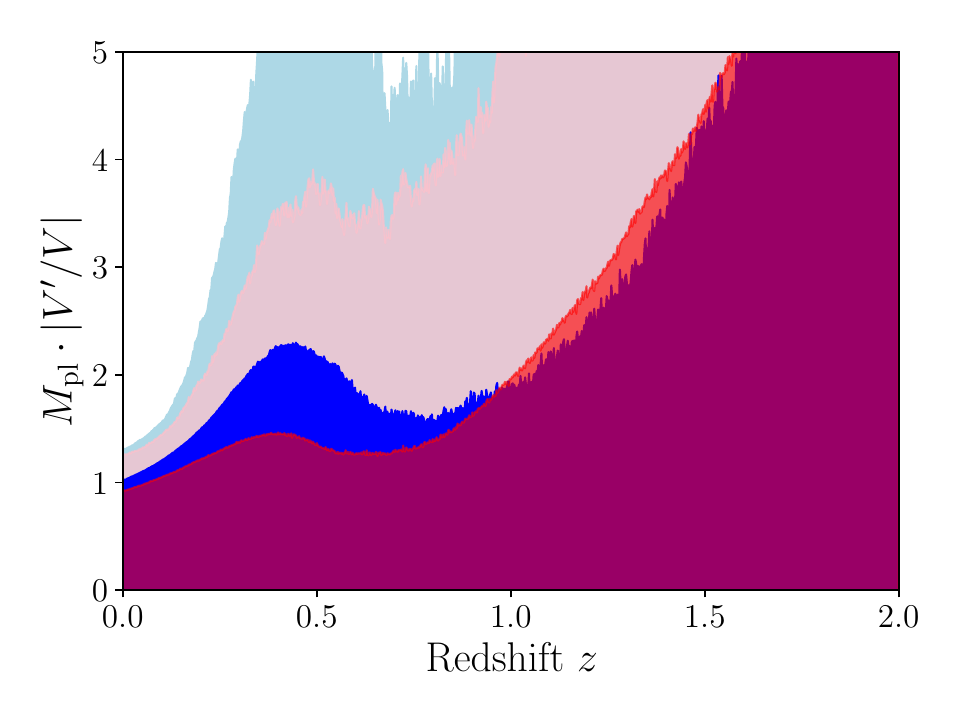}
    \includegraphics[width=0.45\textwidth]{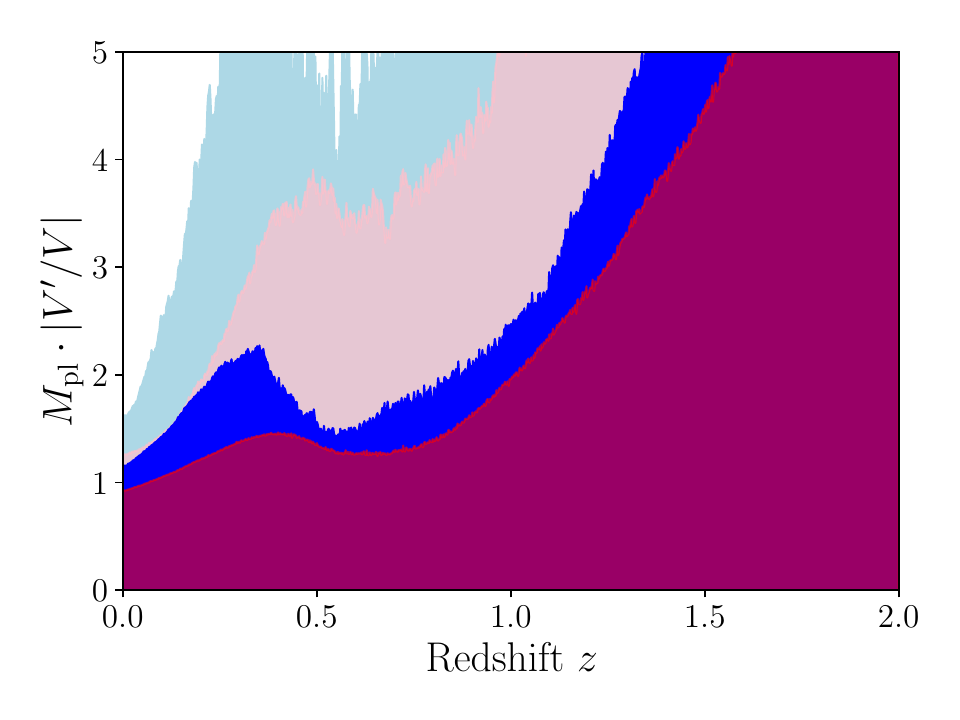}
    \caption{The reconstruction of the first Swampland criterion for a higher $H_0$ (up to $73$km/s/Mpc) or high $\Omega_m$ (up to 0.33). The $68\%$ and 95\% upper quantile bounds are shown in dark blue and light blue, and the corresponding original bounds for comparison (from \cref{fig:reconstr_model_indep}, right panel) are shown in slightly transparent red and pink.}
    \label{fig:reconstr_model_varied}
\end{figure}

Finally, the remaining histories can be used for bounds on $\lambda(z)$. However, one last pitfall remains when taking the absolute value of the reconstructed $V'/V$ (which typically has large contributions from negative values). First, it is not entirely consistent to cite mean and standard deviation for $\lambda(z)$, especially after taking an absolute value. We explicitly show the non-Gaussianity in both cases in \cref{fig:reconstr_nongaussian}. Instead, one can use quantiles, though here too some care should be used. In principle, to put an upper bound on $|V'/V|$, only upper limit quantiles (such as $68\%$ or $95\%$) should be used. We show the results of such an analysis in \cref{fig:reconstr_model_indep}, which agree well with those from \cite[Fig.~1]{Yang2020} and \cite[Fig.~3]{Elizalde2019}. However, this is not always done, as \cite{Elizalde2019,Yang2020,Arjona2021} all use lower bounds as well. Here some careful consideration has to be made with the absolute value transformation, as otherwise interpretation is tricky. The naive lower bound of simply taking the lower quantile (e.g. 2.5\% for a two-sided $2\sigma$ limit) leads to obviously wrong conclusions, since by definition it is positive definite and greater than zero even if the original function is consistent with zero at the $2\sigma$ limit. We compare in \cref{fig:reconstr_quantiles} different possible ways to define meaningful bounds for the absolute value. In the left panel of \cref{fig:reconstr_quantiles} we show the quantiles corresponding to the $1\sigma$ and $2\sigma$ limits for $V'/V$ without absolute value, and simply transpose these into $|V'/V|$ in the right panel by taking $-V'/V$ and discarding any of the positive values (since those would give smaller bounds). As a comparison, we also display the true one-sided upper limits at the $1\sigma$ and $2\sigma$ level, which we also give in \cref{fig:reconstr_model_indep,fig:reconstr_model_varied}. Finally, we show the upper limits using the half-sided probabilities ($84.13\%$ and $97.725\%$) in dark and light green as well. Obviously the simplest treatment of simply cutting/transposing the graph does give decent results, though these cannot be interpreted directly as one-sided upper limits, even when the curve touches the axis. On the other hand, the one-sided limits trivially cannot show the interesting slight favor for non-zero $V'/V$ present at $z=0$. 
\begin{figure}
    \centering
    \includegraphics[width=0.7\textwidth]{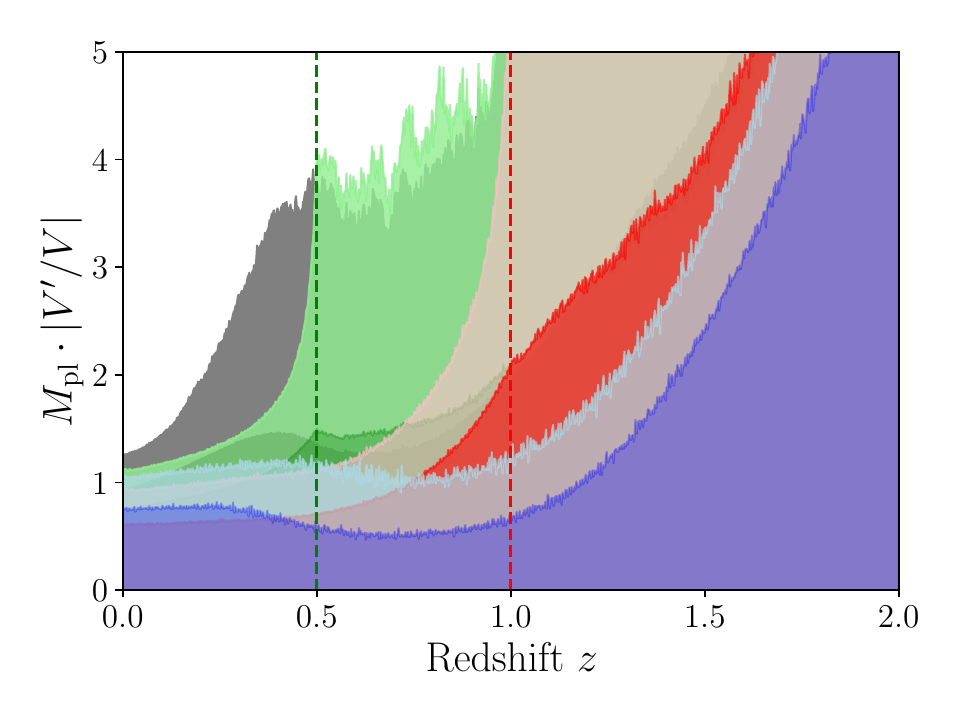}
    \caption{Different reconstructions of the first Swampland criterion with different requirements for validity (see text). The black reconstruction in the background is the one from \cref{fig:reconstr_model_indep} (right panel). Note the different $x$-axis range. The green reconstruction discards all histories, which have any non-physicality at $z<0.5$ (green dashed line), the red reconstruction discards all histories which have any non-physicality at $z<1$ (red dashed line), while the blue reconstruction discards all histories which have any non-physicality in the entire redshift range $z \in [0,2]$. Note the similarity of the blue reconstruction to \cite[Fig.~2]{Arjona2021} and the improved similarity of the green reconstruction to \cite[Fig.~3]{Elizalde2019}.}
    \label{fig:reconstr_validity}
\end{figure}

Coming back to the other pitfalls, we show in \cref{fig:reconstr_model_varied} the dependence on the fiducial constraint on the assumed value of $H_0$ and $\Omega_m$, showing that there is a rather non-negligible dependence that should be accounted for when citing any constraints. Indeed, a higher value of $H_0$ or $\Omega_m$ can easily weaken the constraints by more than 50\% at $z\sim 0.4$.

However, by far the biggest impact is from how to treat non-physical regions. So far, we have always simply ignored the individual samples with unphysical density or field speed. However, this leads to a different number of remaining samples for each redshift. If, instead, we discard a complete history if at any redshift sample there is a negative value of $\rho_\phi$ or its derivative (unphysicality), we instead get the constraints displayed in \cref{fig:reconstr_validity} (blue reconstruction). This reconstruction is more constraining than the original reconstruction, since histories which have larger variations of $V'/V$ also typically feature larger $H'$ or $H''$, which gives a more varying or curved $H(z)$, and thus a higher probability for that history to also include regions of unphysicality. In \cref{fig:reconstr_model_varied} we explicitly see that the bump of less constraining power at $z \sim 0.4$ is at least partially caused by the varying count of the considered histories as a function of redshift, as it does not occur in any of the reconstructions which discard samples if there is an unphysicality at any redshift for the given history. We also show how the reconstruction changes by changing the redshift range over which this condition of physicality is imposed. Importantly, it remains to be stressed that even the reconstruction that discards any unphysical histories for redshift range [0,2] could be made even more stringent by considering an even broader redshift range over which to require physicality. However, since the data only reach up until $z \simeq 2$, we decided not to push the requirement of physicality further. 

In summary, there are many pitfalls to the model-independent reconstruction, and the analysis is certainly not without caveats. The limits of any such analysis should be considered only in the context of these analysis choices, as otherwise conclusions are possibly severely biased, as we have shown in this section. These pitfalls include (non-exhaustive list):
\enlargethispage*{1\baselineskip}
\begin{itemize}
    \item A possibly naive treatment of cosmic chronometer data systematics
    \item Rescaling of supernovae of BAO data using some fiducial cosmology
    \item Requirement to take two or three derivatives, with assumptions about functional smoothness
    \item Using a fiducial model for the other components (e.g.~CDM+baryons)
    \item Regions of non-physicality have to be explicitly removed
    \item Taking the absolute value and how corresponding bounds are derived
\end{itemize}

\subsection{Forecast constraints}
\begin{figure}
    \centering
    \includegraphics[width=0.8\textwidth]{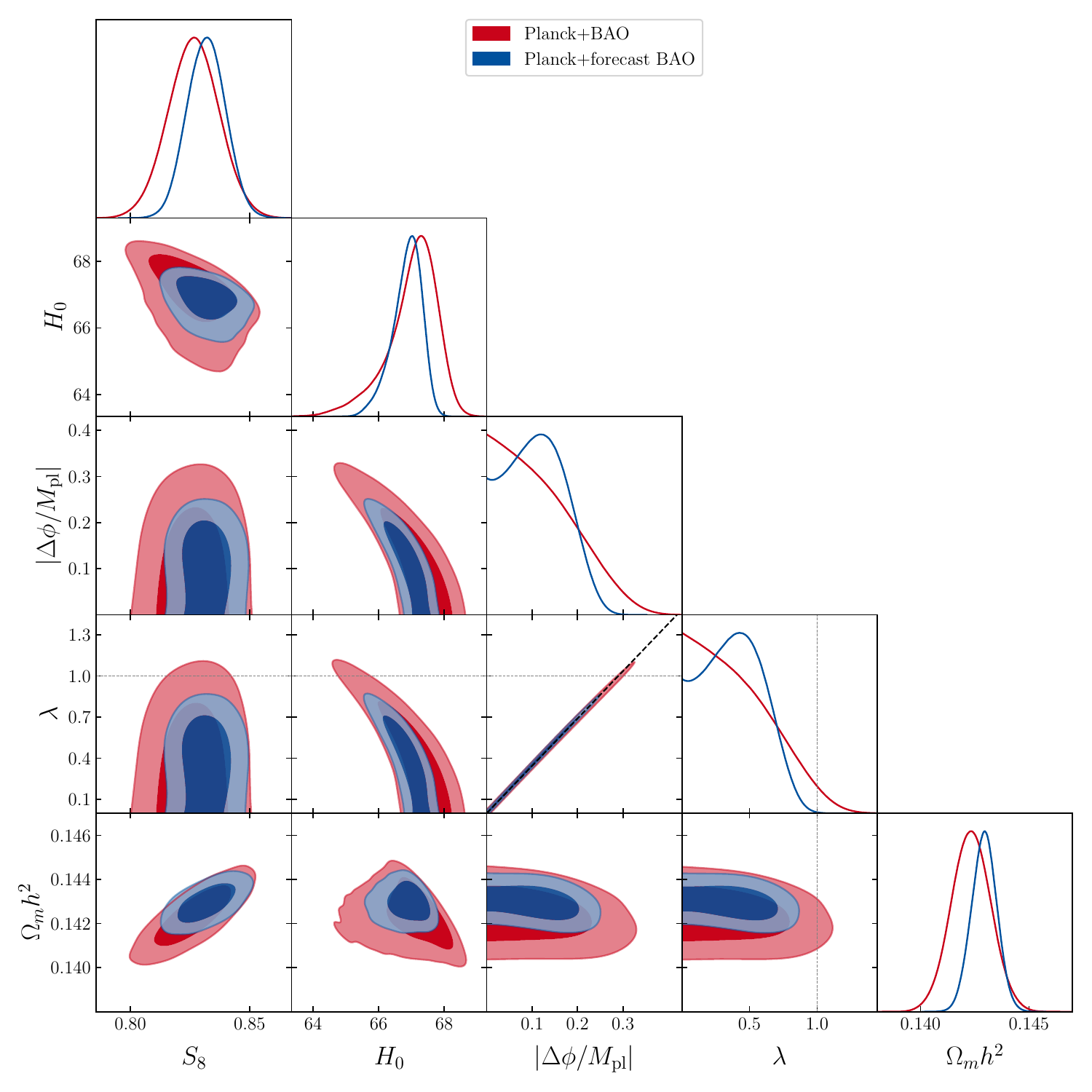}
    \caption{Same as \cref{fig:cqf_summary}, but instead adding forecasts for DESI.}
    \label{fig:cqf_desi}
\end{figure}
We also show the forecast constraints on the exponential case of \cref{sssec:exponential} from early forecasts of the DESI experiment error bars from \cite[Sec.~2.4.3]{DESI2016}, where we use the slightly optimistic 14\,000 deg${}^{2}$ assumptions, including the use of Lyman-$\alpha$ data (Tab. 2.3 and 2.7). A more detailed analysis using more realistic DESI early data releases is outside the scope of this work. The resulting constraints are shown in \cref{fig:cqf_desi}. There is mild improvement on the constraining power of $\lambda$ visible, with constraints improving from $\lambda < 0.94$ to $\lambda < 0.75$ at 95\% CL, with the corresponding probabilities to exceed unity decreasing from $2.9\%$ to $\sim 7 \cdot 10^{-5}$. This 25\% improvement of the constraint would make this the most precise measurement of $\lambda$ to date (apart from artificial choices, such as using old Pantheon data or imposing a prior on $H_0$, which is incompatible with the other datasets such as Planck).

\section{Conclusions}\label{sec:conclusions}

We have summarized and presented the current status of the Swampland conjectures in terms of their impact on late-time cosmology. Already from analytical arguments it is clear that there must be some mild degree of tension between the de Sitter conjectures requiring large potential slopes or curvatures and the fact that the universe is observed to undergo an accelerated expansion. We confirm this argument with current data, showing that indeed in most viable quintessence models of dark energy these conjectures are at a mild level of tension. However, surprisingly we find that this tension did not strongly increase when adding newer BAO and Pantheon data, instead even decreasing. We argued that this is a consequence of the slightly lower dark energy abundance preferred by these probes, which allows for more flexibility in its precise evolution. In the exponential potential case, we find similar constraints to previous works \cite{Raveri2019}, leading to a constraint of $\lambda = \mpl |V'/V| < 0.85$ (95\% CL) when using all currently available data, and $\lambda < 0.60$ (95\% CL) when adding a prior on $H_0$ from \cite{Riess2022}, though this addition of a prior in high tension with the other datasets should be taken with a grain of salt. When instead using older BAO and supernovae data, this bound can in principle be increased up to $\lambda < 0.46$ (95\% CL). We generally find that in potentials where $\lambda(z)$ is not constant that the constraints are tighter at low redshift $z \lesssim 2$, while being slightly more permissive at high redshift where the dark energy becomes irrelevant. With a view towards the second de Sitter criterion on $g(z) = - \mpl^2 V''/V$, we find that the cosine potential provides similar constraints on $\lambda(z)$, though large values of $g(z)$ are still allowed in a part of the parameter space.
Overall, in all cases of specific potentials we investigated, we find that there is a mild tension on the de Sitter Swampland criterion, while the distance criterion (requiring no trans-Planckian field excursions) is typically well fulfilled in all cases. However, the data is not yet precise enough to exclude order-unity slopes or large values of the second Swampland criterion.

We also outlined that there is a high degree of fine-tuning necessary to keep the coupling between the quintessence field and the electromagnetic sector tiny (or one has to invoke some deeper theoretical argument that prevents such interactions). Indeed, with the newest atomic clock likelihoods we find a constraint on the interaction as $|\zeta|<1.5 \cdot 10^{-7}$ (95\% CL), which very strongly excludes any kind of `natural' $\mathcal{O}(0.1-10)$ coupling. As such, this fine-tuning argument presents a rather steep theoretical hurdle for consistent unified theories.

Finally, we argue that model-independent reconstructions of the Swampland criteria that have been frequently presented in the literature are subject to some pitfalls that need to be avoided for proper conclusions from the given analysis. We also argue that two of these pitfalls in particular dominate the overall conclusions from such an analysis. First, the given emulation of the expansion history has to be performed with a given modeling of the smoothness of said history, as otherwise the required derivatives cannot be taken. Second, the enforcement of a physically positive kinetic term and positive field energy density are non-trivial, and depending on the adopted method the results can change quite drastically. Together these problems imply that one can only draw conclusions from the model-independent reconstruction in the context of these major assumptions, and different assumptions can lead to different conclusions.

Overall, we find that current data are only in a relatively mild tension with the Swampland criteria, still giving string theorists large room for model building. This picture could change, however, with future data that will constrain the expansion history to an ever-increasing degree of precision. In particular, we show that forecasted constraints from DESI \cite{DESI2016} would allow for constraints of $\lambda < 0.75$ in the late universe (already improving over the combination of all currently available data, except for the incompatible $H_0$ prior). As such, it might be only a matter of time until the de-Sitter Swampland criteria become incompatible with cosmological observations.

\section*{Acknowledgements}
%\begin{acknowledgements}
N.~S.~acknowledges support from the Maria de Maetzu fellowship grant: CEX2019-000918-M, financiado por MCIN/AEI/10.13039/501100011033. This work was financed by Portuguese funds through FCT - Funda\c c\~ao para a Ci\^encia e a Tecnologia in the framework of the project 2022.04048.PTDC. JDFD is supported by an FCT fellowship, grant number SFRH/BD/150990/2021. CJM also acknowledges FCT and POCH/FSE (EC) support through the grant titled Investigador FCT Contract 2021.01214.CEECIND/CP1658/CT0001.

%\end{acknowledgements}
\bibliography{biblio}
\bibliographystyle{JHEP}

\appendix
\end{document}